\documentclass[twocolumn,showpacs,preprintnumbers,amsmath,amssymb]{revtex4}
\pdfoutput=1
\usepackage{graphicx}
\usepackage{dcolumn}
\usepackage{bm}

\usepackage{amsmath}
\usepackage{amssymb}
\usepackage{pxfonts}
\usepackage{graphicx}
\usepackage{eucal}
\usepackage{mathrsfs}
\usepackage{pifont}
\usepackage[usenames]{color}

\newcommand{\brk}[1]{\left(#1\right)}          
\newcommand{\Brk}[1]{\left[#1\right]}          
\newcommand{\BRK}[1]{\left\{#1\right\}}        
\newcommand{\mymat}[1]{\begin{pmatrix} #1 \end{pmatrix}}

\newcommand{\deriv}[2]{\frac{d#1}{d#2}}

\newcommand{\figref}[1]{Figure~\ref{#1}}

\newcommand{\tabref}[1]{Table~\ref{#1}}

\newcommand{\beq}{\begin{equation}}
\newcommand{\eeq}{\end{equation}}

\newcommand{\bs}[1]{\boldsymbol{#1}}
\providecommand{\half}{\frac{1}{2}}

\newcommand{\R}{\bbR}

\newcommand{\Textand}{\qquad\text{ and }\qquad}

\newcommand{\putfig}[2]{
   \medskip\centerline{\includegraphics[height=#1in]{#2}}\medskip
}

\newcommand{\calA}{{\mathcal{A}}}
\newcommand{\calH}{{\mathcal{H}}}
\newcommand{\calD}{{\mathcal{D}}}
\newcommand{\scrF}{{\mathscr{F}}}
\newcommand{\calN}{{\mathcal{N}}}
\newcommand{\calO}{{\mathcal{O}}}
\newcommand{\calS}{{\mathcal{S}}}
\newcommand{\bbR}{{\mathbb{R}}}

\renewcommand{\r}{\boldsymbol{r}}
\newcommand{\x}{\boldsymbol{x}}
\newcommand{\f}{\boldsymbol{f}}
\newcommand{\go}{\bar{g}}
\newcommand{\e}{\epsilon}
\renewcommand{\a}{\alpha}
\renewcommand{\b}{\beta}
\newcommand{\g}{\gamma}
\renewcommand{\d}{\delta}
\renewcommand{\O}{\calO}
\newcommand{\Chr}[3]{\Gamma^{#1}_{#2#3}}
\newcommand{\rmin}{r_{\text{min}}}
\newcommand{\rmax}{r_{\text{max}}}
\newcommand{\N}{\boldsymbol{\calN}}
\newcommand{\tw}{\tilde{w}}
\newcommand{\tv}{\tilde{v}}
\newcommand{\SurfaceElement}{dS}
\newcommand{\strain}{\varepsilon}
\newcommand{\stress}{s}
\newcommand{\thickness}{t}
\newcommand{\tc}{\thickness_b}
\newcommand{\tco}{\thickness_{\text{c.o.}}}
\newcommand{\Ft}{F_\thickness}
\newcommand{\ft}{\f_\thickness}
\newcommand{\ellBL}{\ell_{\text{B.L.}}}
\newcommand{\df}[1]{\partial_{#1}\f}
\newcommand{\ddf}[2]{\partial_{#2}\partial_{#1}\f}

\newcommand{\tr}{\operatorname{tr}}
\newcommand{\dist}{\operatorname{dist}}

\newcommand{\figsize}{1.25}

\newcommand{\parag}[1]{\bigskip\noindent {\bfseries #1}\hspace{0.5cm}}

\begin{document}

\preprint{PrePrint}

\title{Buckling transition and boundary layer in non-Euclidean plates}

\author{Efi Efrati,}
\author{Eran Sharon,}
\affiliation{The Racah Institute of Physics, The Hebrew University, Jerusalem 91904, Israel}
\author{Raz Kupferman}
\affiliation{Institute of Mathematics, The Hebrew University, Jerusalem 91904, Israel}

\date{\today}

\begin{abstract}
Non-Euclidean plates are thin elastic bodies having no stress-free configuration, hence exhibiting residual stresses in the absence of external constraints. These bodies are endowed with a three-dimensional reference metric, which may not necessarily be immersible in physical space.  Here, based on a recently developed theory for such bodies,  we characterize the transition from flat to buckled equilibrium configurations at a critical value of the plate thickness. Depending of the reference metric, the buckling transition may be either continuous or discontinuous.
In the infinitely thin plate limit, under the assumption that a limiting configuration exists, we show that the limit is a configuration that minimizes the bending content, amongst all configurations with zero stretching content (isometric immersions of the mid-surface). For small but finite plate thickness we show the formation of a boundary layer, whose size scales with the square root of the plate thickness, and whose shape is determined by a balance between stretching and bending energies.
\end{abstract}



\maketitle


\section{Introduction}

The classical literature on thin elastic bodies deals primarily with two types of bodies---\emph{plates} and \emph{shells}. Mathematically, a plate can be viewed as a continuous stack of identical flat surfaces glued together, whereas a shell can be viewed as a continuous stack of non-identical (and not necessarily flat) surfaces glued together.
The term non-Euclidean plate was coined in \cite{ESK08} to describe thin elastic bodies which, like plates, do not exhibit structural variations across  their thin dimension, and yet, unlike plates, do not have a planar rest configuration. Such elastic bodies can neither be described as shells, which bear structural variations across their thin dimension (e.g., shells do not display reflectional symmetry about the mid-surface), and possess curved stress-free rest configurations.
Non-Euclidean plates exhibit residual stresses even in the absence of external constraints, and are therefore inherently frustrated.

\begin{figure}
\putfig{2}{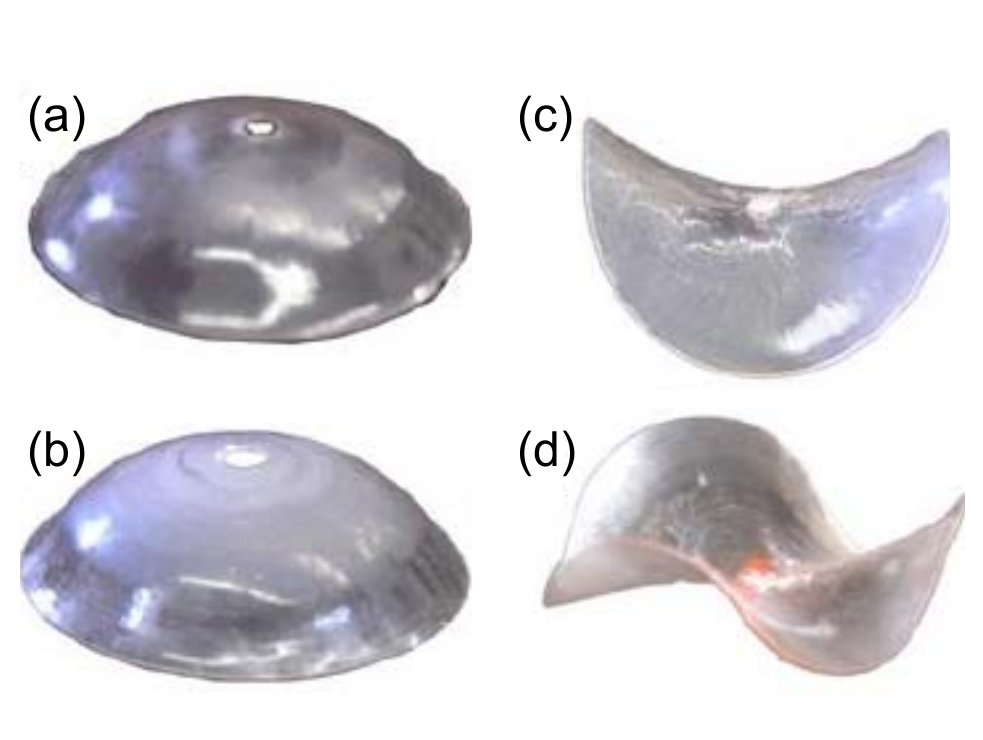}
\caption{(Color online) Four elastic plates made of thermo-responsive gel as described in \cite{KES07}. All four structures bear no structural variation across their thickness. Their  radius is 3 cm.  The mid-surface of the positively curved discs (a and b) possess a reference metric of constant Gaussian curvature $K=0.11$ cm$^{-2}$ . The mid-surface of the negatively curved surfaces (c and d) possess a reference metric of constant Gaussian curvature of opposite sign
$K= - 0.11$ cm$^{-2} $. Plates (a) and (c) are 0.75 mm thick, whereas plates (b) and (d) are 0.6 mm thick.   }
\label{fig:gels}
\end{figure}

Elastic bodies having such properties are ubiquitous in biology. Growing tissues, such as plant leaves,  are relatively thin elastic structures that may exhibit complex equilibrium configurations in the absence of external forces \cite{NCCC03,SRS07}. In addition, thin elastic bodies that have no stress-free configuration have been engineered in a laboratory \cite{KES07}, for example, the environmentally responsive gels shown in \figref{fig:gels}.

There are various ways to treat elastic bodies which exhibit residual stress. One way is to treat the residual stress as a physical field and characterize its properties \cite{Hog85}. Another is to decompose the deformation gradient into a product of a plastic (or growth) process, which deforms the body from a rest configuration into some ``virtual" configuration, and an elastic relaxation from the virtual configuration to the current configuration \cite{BG05,Sid82}.
A third approach is to decompose the strain tensor additively into a plastic (or growth) strain, leading to a virtual configuration, and an elastic strain \cite{SD02,GN71}. The treatment presented here for residually stressed
three-dimensional bodies is very similar to the third approach.
All deformations which are not of elastic nature are completely ignored, i.e. it is assumed that the virtual configuration which the growth or plastic deformation led to is known, and the appropriate elastic relaxation is solved. This in turn enables us to treat large ``plastic strains" in a non-iterative manner. 

A static theory of non-Euclidean plates was developed in \cite{ESK08} following the fundamental principles laid by Truesdell \cite{Tru52} and its modern interpretation by Ciarlet \cite{CG05,Cia05}. The starting point in \cite{ESK08} is the formulation of a covariant three-dimensional elasticity theory in the form of an energy functional. A first notable property of this energy functional is its expression in terms of the three-dimensional metric of the configuration. Specifically, the energy density is quadratic in the deviation of the metric from a reference metric. This deviation of the metric is a strain, which reduces to the standard Green-Saint Venant strain for bodies that have a rest configuration. The second notable property of our model is that the reference metric is not required to have a vanishing Riemannian curvature tensor, i.e., it may not be immersible in $\R^3$ (hence the name ``incompatible elasticity theory"). As a result, there exist no rest configurations in  which the strains vanish everywhere, hence the state of frustration.
In a second step, a two-dimensional elasticity theory is derived, using a generalization of the standard Kirchhoff-Love assumptions \cite{Kir50,Lov27}. The end result is an energy functional which depends on surface properties of the midplane of the plate, namely, on the first and second fundamental forms. Like in the classical F\"oppl-von-K\'arm\'an theory \cite{Kar10} the energy functional is a sum of a stretching term and a bending term.
The bending term is minimized in flat configurations, whereas the stretching term measures, in an $L^2$ sense,  deviations of the 2D surface metric from a prescribed reference metric, and vanishes only in surfaces which are isometric immersions of the given 2D metric.
The lack of immersibility of the three-dimensional metric manifests in the lack of a planar stretching-free configuration.

At this stage we have a new model, which we believe to be applicable to a large variety of physical and biological systems, whose properties are governed by essentially two-dimensional shaping mechanisms.
In  \cite{ESK08}, a single
application  was demonstrated for the case of an unconstrained thin plate, whose two-dimensional reference metric is that of a punctured spherical cap.
A buckling transition was shown to occur at a critical plate thickness.


In this paper we study two behaviors exhibited by unconstrained non-Euclidean plates. First, we study the transitions from flat to buckled equilibrium states as the plate thickness crosses a critical value---the buckling threshold. We derive an explicit expression for the critical thickness in terms of the stress field in the planar configuration. An immediate implication is that the plane-stress solution always becomes unstable for sufficiently thin plates, provided that it is not trivial, i.e., that the stress is not identically zero. We apply this analysis to three reference geometries of constant Gaussian curvature of different types---an elliptic, a flat and a hyperbolic metric. We show that the buckling transition may be either continuous (super-critical) or discontinuous (sub-critical).
In particular, we show that the buckling threshold may deviate significantly from the so-called crossover point, which is based on a balance between the plane-stress energy and the energy that minimizes the Willmore functional. We show an example in which the crossover thickness underestimates the buckling threshold by more than an order of magnitude.

Second, we analyze the equilibrium configurations and energies in the limit where the plate thickness tends to zero.
We show that if a limit configuration exists, then it is the minimizer of the bending content amongst all configurations with zero stretching content, i.e., the Willmore energy minimizer amongst all isometric immersion of  the 2D reference metric \cite{Wil92}. For a small but finite thickness, deviations from isometric immersions are more pronounced near the free boundary of the domain, forming a boundary layer, which we obtain in explicit form. In particular, the size of this boundary layer is found to scale with the square root of the plate thickness.

\section{Theory of non-Euclidean plates}
\label{sec:model}

In this section we briefly review the modeling of non-Euclidean plates, first described in \cite{ESK08}. The starting point is a three-dimensional covariant elasticity theory, based on the principles of hyper-elasticity \cite{Tru52}: the elastic energy is a volume integral over an energy density, which depends only on (i) the local value of the metric tensor and (ii) local characteristics of the material that are independent of the configuration
(the use of the metric tensor, rather than the deformation, as primitive variable has been originally proposed by Antman \cite{Ant76a}, and has been recently advocated by Ciarlet and co-workers \cite{CM04b,CM04,Cia03,CG05,CC04}).

Let $\Omega\subset\R^3$ be an elastic body endowed with a set of material curvilinear coordinates $\x = (x^1,x^2,x^3)\subset \calD\subset\R^3$. Let $\r$ denote the mapping from the domain of parametrization $\calD$ into $\Omega$---$\r(\x)$ is called the configuration---then the induced Euclidean metric is $g_{ij} = \partial_i\r\cdot\partial_j\r$, where $\partial_i=\partial/\partial x^i$.   Our model assumes the existence of a reference metric $\go_{ij}(\x)$, such that the elastic energy density vanishes at a point $\x$ if and only if the actual metric coincides with the reference metric at that point, $g_{ij}(\x) = \go_{ij}(\x)$. While the reference metric is required to satisfy the properties of  a metric---it is symmetric positive-definite---it is not necessarily immersible in $\R^3$, hence the name of the theory as ``incompatible three-dimensional elasticity"

The strain tensor is defined as half the deviation of the metric from the reference metric,
\[
\strain_{ij} = \half(g_{ij} - \go_{ij}).
\]
It coincides with the Green-Saint Venant strain tensor in the case where
there exists a rest configuration, and the curvilinear coordinates form a Cartesian parametrization in the rest configuration, i.e., when $\go_{ij} = \d_{ij}$. For small deviations of the metric from the reference metric, the energy functional is truncated at the first non-trivial term, i.e., it is quadratic in the strain tensor, yielding
\begin{equation}
E = \int_\calD w(g) \sqrt{|\go|}\,dx^1 dx^2 dx^3,
\label{eq:3Denergy}
\end{equation}
where
\[
w(g) = \half A^{ijkl} \strain_{ij} \strain_{kl},
\]
and
\beq
A^{ijlk} = \lambda \go^{ij} \go^{kl} + \mu\brk{\go^{ik} \go^{jl} + \go^{il} \go^{jk}},
\label{eq:Aijkl}
\eeq
where $\lambda,\mu$ are the Lam\'e coefficients.

\parag{Comments:}

\begin{enumerate}
\item We adopt the Einstein summation convention whereby repeated indices imply summation.

\item Latin lowercase characters $i,j,\dots=1,2,3$ are used to denote indices of three-dimensional tensors. We will use below Greek characters $\a,\b,\dots=1,2$  to denote indices of two-dimensional tensors.  For any tensor $a_{ij}$, $|a|$ denotes its determinant.

\item The tensor $\go^{ij}$ is the tensor reciprocal to $\go_{ij}$. The raising and lowering of indices is only defined with respect to the reference metric. For example, the tensor $g^{ij}$ is defined as $\go^{ik} \go^{jl} g_{kl}$ and not as the reciprocal of $g_{ij}$, which we denote by $(g^{-1})^{ij}$.

\item  The volume element in \eqref{eq:3Denergy} is determined by the reference metric rather than the actual metric.
Note that it is not a priori clear whether the volume element should be derived from the reference metric or from the actual metric. In any case, the difference between the two choices is of higher order in the strain.

\item
The structure \eqref{eq:Aijkl} of the elastic tensor is imposed by the assumption of spatial isotropy.

\item
In standard (or ``compatible") nonlinear elasticity, the energy density is sometimes written in terms of the Euclidean distance of the deformation gradient,  $\nabla\r$, from the group of proper rotations, SO(3). In the same spirit, the energy density in an incompatible elasticity theory
may be expressed  in terms of
\[
\dist(\nabla\r,\scrF_{\go}),
\]
where $\scrF_{\go}$ is the set of matrices, $R$, such that $R^T R = \go$.

\item
In summary, given the reference metric $\go$, the elastic problem is formulated as follows: find the metric $g$ that minimizes the energy functional \eqref{eq:3Denergy}, subject to the constraint that it is embeddable, in particular, that the corresponding Riemann curvature tensor vanishes.

\end{enumerate}

With a three-dimensional elasticity theory in hand, we focus the attention on plate-like structures. We define a plate to be an elastic body for which there exists a parametrization in which the reference metric takes the form
\begin{equation}
\go_{ij} = \mymat{\go_{11} & \go_{12} & 0 \\ \go_{21} & \go_{22} & 0 \\ 0 & 0 & 1},
\label{eq:go}
\end{equation}
with $\go_{ij}$ independent of $x^3$.
The plate is called \emph{even} if $\calD = \calS\times[-\thickness/2,\thickness/2]$ with $\calS\subset\R^2$, and $t$ a constant, and \emph{thin} if $\thickness$ is much smaller than all other dimensions.  We identify the two-dimensional tensor $\go_{\a\b}$ as the metric tensor of a surface. By assumption it is constant across the plate thickness. It is easy to see that the three-dimensional reference metric \eqref{eq:go} is immersible in $\R^3$ if and only if the two-dimensional reference metric $\go_{\a\b}$ has zero Gaussian curvature.

To derive a reduced two-dimensional energy density in terms of the mid-surface configuration we used an adaptation of the Kirchhoff-Love assumptions \cite{Kir50,Lov27}: We first assume that the stress is parallel to the mid-surface, and then that $\strain_{i3}=0$ (the order in which the two assumptions are used is essential).
Integrating the energy functional \eqref{eq:3Denergy} over the thin direction, it takes straightforward  manipulations to derive an energy functional which depends on the first and second fundamental forms of the mid-surface. Specifically, let $\f(x^1,x^2) = \r(x^1,x^2,0)$ be the immersion of the mid-surface, then
\[
g_{\a\b} = \df\a\cdot\df\b
\Textand
h_{\a\b} = \ddf\a\b\cdot\N,
\]
are the first and second fundamental forms, where $\N$ is the unit vector normal to the mid-surface. The energy functional is given by
\beq
W =  \thickness\, E_S + \thickness^3\, E_B,
\label{eq:Wfunc}
\eeq
where
\[
E_S = \int_\calS w_S \,\SurfaceElement
\qquad
E_B = \int_\calS w_B \,\SurfaceElement
\]
are called the stretching and bending contents,
\beq
\begin{aligned}
w_S &= \frac{1}{8} \calA^{\a\b\g\d}(g_{\a\b} - \go_{\a\b})
(g_{\g\d} - \go_{\g\d}) \\
w_B &=  \frac{1}{24} \calA^{\a\b\g\d} h_{\a\b} h_{\g\d}
\end{aligned}
\label{eq:wswb}
\eeq
are their respective densities, where
\[
\calA^{\a\b\g\d} = \frac{Y}{2(1+\nu)}\brk{\frac{2\nu}{1-\nu} \go^{\a\b}\go^{\g\d}
+\go^{\a\g}\go^{\b\d} + \go^{\a\d}\go^{\b\g}},
\]
and $\SurfaceElement=\sqrt{|\go|}dx^1 dx^2$ is the infinitesimal surface element.
The coefficients $Y$ and $\nu$ are Young's modulus and the Poisson ratio, which can be related to the Lam\'e coefficients. The elastic energy is positive definite for constant values of $Y>0$ and $-1\le\nu<\half$.

\parag{Comments:}
\begin{enumerate}
\item Like in the classical F\"oppl-von-K\'arm\'an and Koiter theories, the energy functional is a sum of (i) a stretching energy, which scales with the plate thickness and attains a minimum for an isometric immersion of the mid-plane surface, and (ii) a bending energy, which scales like the third power of the thickness  and attains a minimum for flat configurations. The equilibrium configuration is the minimizer of the sum of both stretching and bending energies.

\item
Rather than working with the energy $W$ given by \eqref{eq:Wfunc} we will work with the energy-per-unit-thickness,
\beq
E = \frac{W}{\thickness} = E_S + \thickness^2 E_B.
\label{eq:Efunc}
\eeq
Henceforth, we will refer to $E$ as ``the energy". Thus
the stretching energy is $\thickness$-independent whereas the bending energy scales with $\thickness^2$.
Obviously, both $W$ and $E$ have the same minimizer. In addition, we rescale the units of energy by a factor of $Y/(1+\nu)$, such that the tensor $\calA^{\a\b\g\d}$ takes the final form
\[
\calA^{\a\b\g\d} = \frac{\nu}{1-\nu} \go^{\a\b}\go^{\g\d}
+ \half\brk{\go^{\a\g}\go^{\b\d} + \go^{\a\d}\go^{\b\g}}.
\]

\item
A different derivation of an elastic functional similar to \eqref{eq:Efunc} may be found in \cite{Cia05}. Eq. \eqref{eq:Efunc} may be identified as the elastic energy in the Koiter shell model when the ``target bending tensor" $\bar{h}_{\a\b}$ is set to zero \cite{Koi66}.

\item
With the above rescaling the stretching and bending density contents can be written in the more compact form
\[
\begin{aligned}
w_S &= \frac{\nu}{8(1-\nu)} [\tr (\go^{-1} g - I)]^2 + \frac{1}{8} \tr[ (\go^{-1} g - I)^2] \\
w_B &= \frac{\nu}{24(1-\nu)} [\tr (\go^{-1} h)]^2 + \frac{1}{24} \tr[ (\go^{-1} h)^2].
\end{aligned}
\]

\item
The energy functional is expressed in terms of the first two fundamental forms $g_{\a\b}$ and $h_{\a\b}$ of the mid-surface. The two forms are not independent: they must satisfy the three Gauss-Mainardi-Codazzi compatibility conditions,
\[
\begin{gathered}
\partial_\e h_{\a\b}  - \partial_\b h_{\a\e} =
\Chr\g\a\e h_{\g\b} -
\Chr\g\a\b h_{\g\e} \\
h_{\a\e} h_{\b\eta} -
h_{\a\b} h_{\e\eta}
= g_{\d\eta}
\brk{
\partial_\b \Chr\d\a\e  -
\partial_\e \Chr\d\a\b +
\Chr\g\a\e \Chr\d\g\b -
\Chr\g\a\b \Chr\d\g\e},
\end{gathered}
\]
where the Christoffel symbols are given by
\[
\Chr\g\a\b = \half (g^{-1})^{\g\d}
\brk{\partial_\a g_{\b\d} +
\partial_\b g_{\a\d} - \partial_\d g_{\a\b}}.
\]

\item The two-dimensional stress and moment tensors are defined as
\[
\stress^{\a\b} =  \calA^{\a\b\g\d} \strain_{\g\d}
\Textand
m^{\a\b} = \frac{\thickness^2}{12} \calA^{\a\b\g\d} h_{\g\d},
\]
so that
\[
w_S + \thickness^2 w_B = \half \stress^{\a\b} \strain_{\a\b} + \half m^{\a\b}  h_{\a\b}.
\]

\item A surface $\f(x^1,x^2)$ will be called an isometric immersion if the two-dimensional metric $g_{\a\b}$ coincides with the two-dimensional reference metric $\go_{\a\b}$, i.e., if the stretching energy is zero. In the case of an isometric immersion the  bending content density $w_B$ can be identified with the density of the Willmore functional,
\[
w_W = \frac{1}{24} \brk{\frac{4H^2}{1-\nu} + 2K},
\]
where $H$ and  $K$ are the mean and Gaussian curvatures of the surface \cite{Wil92}.
Note that since $K$ is an isometric invariant, its value is prescribed by the reference metric. 

\item
In summary, the two-dimensional elastic problem is defined as follows: given the two-dimensional reference metric $\go_{\a\b}$, find a symmetric positive definite tensor field $g_{\a\b}$, and a symmetric tensor field $h_{\a\b}$, that together minimize the energy functional \eqref{eq:Efunc}, subject to the constraint that the Gauss-Mainardi-Codazzi equations are satisfied.

\end{enumerate}

\section{The infinitely-thin plate limit}
\label{sec:limit}

In many applications, the elastic body is thin to an extent that the equilibrium configuration (of its mid-surface) remains practically unchanged upon further thinning. In other words, we identify an asymptotic regime, which we may call the infinitely-thin plate limit, which in our model corresponds to the limit $\thickness\to0$. It is important to stress that there are also opposite cases, where the thinner the body is, the more convoluted the equilibrium configuration is, with no evidence that a $\thickness\to0$ limit exists (e.g., in \cite{SMS04} a torn plastic sheet exhibits a self-similar shape, whose cut-off scale is comparable to the thickness of the sheet).

Under the assumption that $\go$ admits an isometric embedding of finite bending content, and that a $\thickness\to0$ (weak) limit configuration exists, we show in Appendix~\ref{app:gamma} that the limit is a minimizer of the Willmore functional amongst all isometric embeddings.

The first assumption that the bending content is finite is non-trivial. If it does not hold, then a limit configuration may not exist. 
We expect, however, the second assumption, regarding the existence of a (weak) limit, to become eventually superfluous, 
yet further analysis is required before this assumption can be relaxed.

\section{The buckling transition}

\subsection{Plane-stress solution}

A configuration is a flat surface if $h_{\a\b}=0$. The configuration $\f(x^1,x^2)$ that minimizes the elastic energy under the constraint that the surface be flat is called the \emph{plane-stress solution}. It is the minimizer of the stretching content, which is given by
\[
E = \half \int_\calS \stress^{\a\b} \strain_{\a\b} \, \SurfaceElement =
\frac{1}{8} \int_\calS \calA^{\a\b\g\d}(g_{\a\b} - \go_{\a\b})(g_{\g\d} - \go_{\g\d})\, \SurfaceElement,
\]
with respect to all flat metrics $g_{\a\b}$.
To find the minimizer, we consider an in-plane perturbation of the surface,
\[
\f \mapsto \f + v^\g \,\df\g.
\]
The reason we perturb the configuration rather than the metric is that the three components of the configuration are independent, whereas the three entries of the metric tensor are constrained by the Gauss-Mainardi-Codazzi relations.
The corresponding variation of the metric is
\[
\d g_{\a\b} = \df\a \cdot \partial_\b(v^\g \df{\g}) + \df\b \cdot\partial_\a(v^\g \df{\g}) + \O(v^2).
\]
Using the fact that $\ddf\a\b \cdot\df\g = \Chr\eta\a\b g_{\eta\g}$, the energy variation is
\[
\delta E = \half \int_\calS \stress^{\a\b} \Brk{g_{\a\g}(\partial_\b v^\g) + \Chr\eta\b\g g_{\eta\a} v^\g}
\, \SurfaceElement + \O(v^2).
\]
Integrating by parts, requiring the first variation to vanish for any perturbation $v^\g$ (and using the identity $\partial_\a g_{\b\g} = \Chr\eta\a\b g_{\eta\g} - \Chr\eta\a\g g_{\eta\b}$), the Euler-Lagrange equations are
\begin{equation}
\frac{1}{\sqrt{|\go|}} \partial_\b\brk{\sqrt{|\go|} \stress^{\eta\b}} +  \Chr\eta\a\b  \stress^{\a\b} = 0,
\label{eq:plane_stress}
\end{equation}
with boundary conditions $\stress^{\a\b} n_\b=0$, where $n_\b$ is the outward unit vector tangent to the plate and normal to its boundary. We refer to \eqref{eq:plane_stress} as the \emph{plane-stress membrane equations}.  Note that the plane-stress equations do not depend on the plate thickness $\thickness$, which only comes into play when there is a competition between stretching and bending energies.

\parag{Comment}
Equation \eqref{eq:plane_stress} is the Euler-Lagrange equation associated with the energy functional \eqref{eq:Efunc} when $\thickness=0$. It is expected to hold  in the limit $\thickness\to0$ even for non-flat  configurations. In general,  \eqref{eq:plane_stress} constitutes two equations for three unknown functions (the three components of the metric tensor $g_{\a\b}$), i.e., the system is under-determined. In the present case, the third equation which removes this under-determinacy is that the  curvatures be identically zero.

\parag{Examples}
In the following examples we consider for simplicity the case of a vanishing poisson ratio ($\nu=0$), hence  $\calA^{\a\b\g\d} = \half(\go^{\a\g} \go^{\b\d}+\go^{\a\d} \go^{\b\g})$.
Denote $x^1=r$ and $x^2=\theta$ and consider a reference metric in semi-geodesic parametrization of the form,
\[
\go_{\a\b}(r,\theta) = \mymat{1 & 0 \\ 0 & \Phi^2(r)},
\]
with $\Phi(r)$ yet to be specified. The domain of parametrization is
\[
(r,\theta) \in [\rmin,\rmax]\times[0,2\pi),
\]
with periodicity in the $\theta$-axis, so that the topology of the body is that of a punctured disc.

The equilibrium configuration is expected to preserve the axisymmetry of the intrinsic geometry, hence we seek plane-stress solutions of the form
\[
\f(r,\theta) = (\phi(r)\,\cos\theta,\phi(r)\,\sin\theta,0).
\]
Elementary calculations show that the resulting two-dimensional metric is
\[
g_{\a\b}  = \mymat{(\phi')^2 & 0 \\ 0 & \phi^2},
\]
where $\phi' = d\phi/dr$,
from which we derive the two-dimensional stress tensor,
\beq
\stress^{\a\b} = \half \go^{\a\g} (g_{\g\d} - \go_{\g\d}) \go^{\d\b} =
\half \mymat{(\phi')^2-1 & 0 \\ 0 & (\phi^2/\Phi^2-1)/\Phi^2}.
\label{eq:PSstress}
\end{equation}
Finally, the Christoffel symbols are given by
\beq
\Chr{r}\a\b = \mymat{\phi''/\phi' & 0 \\ 0 & -\phi/\phi'}
\Textand
\Chr\theta\a\b = \mymat{0 & \phi'/\phi \\ \phi'/\phi & 0},
\label{eq:Chr_example}
\eeq
hence the resulting plane-stress equation is
\begin{equation}
\deriv{}{r}\brk{\Phi \phi'[(\phi')^2-1]} = \frac{\phi}{\Phi}\brk{\frac{\phi^2}{\Phi^2}-1},
\label{eq:plane-stress-example}
\end{equation}
with boundary conditions $\phi'(\rmin) = \phi'(\rmax)=1$.

We solve the plane-stress equation \eqref{eq:plane-stress-example} for three families of metrics:
\begin{enumerate}
\item A family of elliptic metrics,
\begin{equation}
\Phi(r) = \frac{1}{\sqrt{K}} \sin \sqrt{K} r,
\label{eq:elliptic}
\end{equation}
where $K>0$ is the constant Gaussian curvature of the reference metric. Although such a metric is consistent with an infinite set of immersions, the immersion that minimizes the Willmore functional is easily identified---it is a (punctured) spherical cap.

\item A family of conical flat metrics,
\begin{equation}
\Phi(r) = \alpha r,
\label{eq:flat}
\end{equation}
with $\alpha<1$.
Here the isometric immersion that minimizes the Willmore functional has the  form of a truncated cone (a circular frustum).
Note that although the reference metric is flat, all isometric immersions have non-zero bending energy due to the topological constraint (periodicity in the $\theta$ axis).

\item A family of hyperbolic metrics,
\begin{equation}
\Phi(r) = \frac{1}{\sqrt{-K}} \sinh \sqrt{-K} r,
\label{eq:hyperbolic}
\end{equation}
where $K<0$ is the constant Gaussian curvature.
Unlike the two former cases, the minimizer of the Willmore functional amongst all isometric embeddings  is not known explicitly, yet, it is known that isometric embeddings with finite bending content do exist \cite{PS96}.

\end{enumerate}

\begin{figure}
\begin{center}
\includegraphics[height=\figsize in]{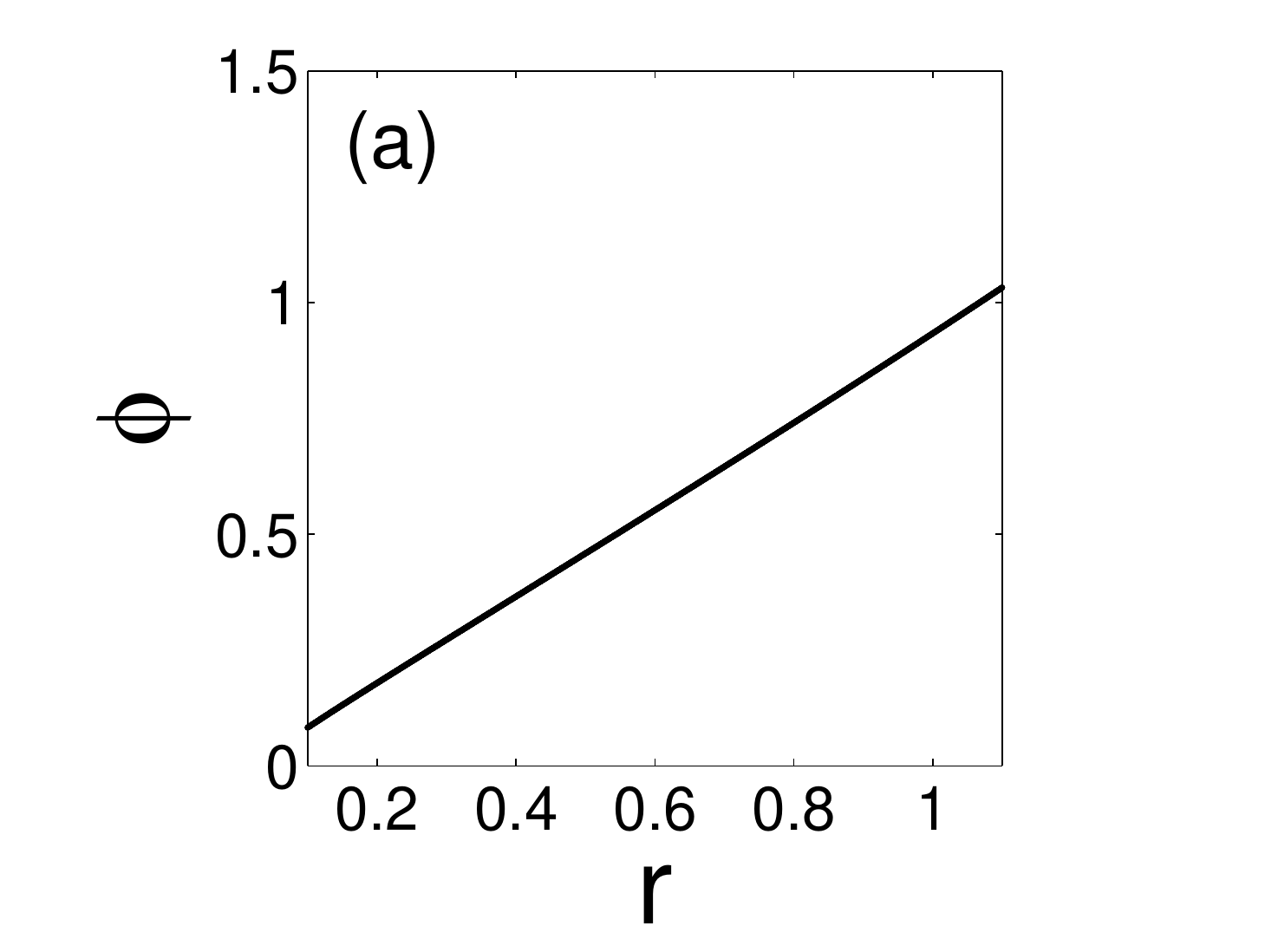}
\includegraphics[height=\figsize in]{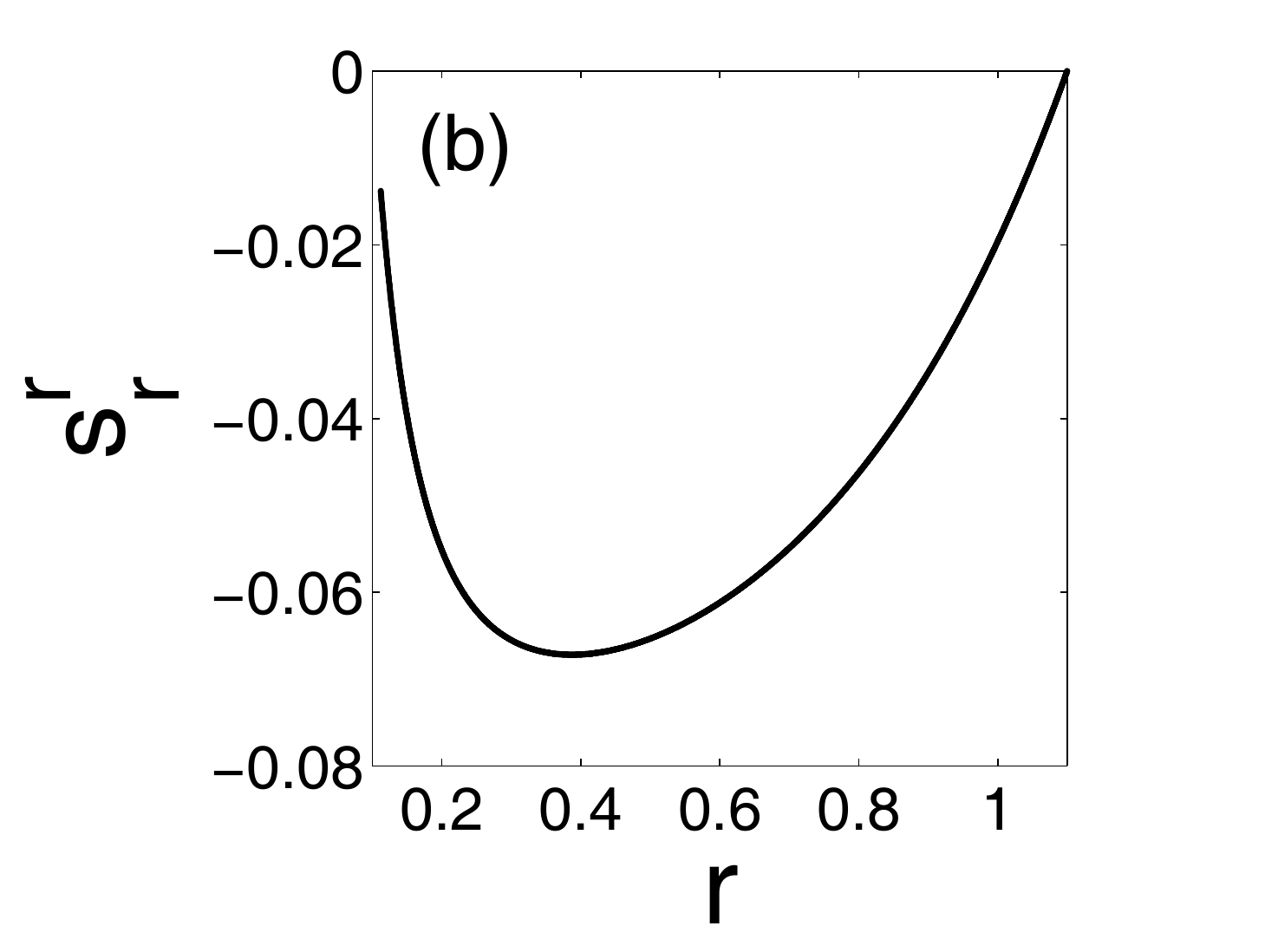}
\includegraphics[height=\figsize in]{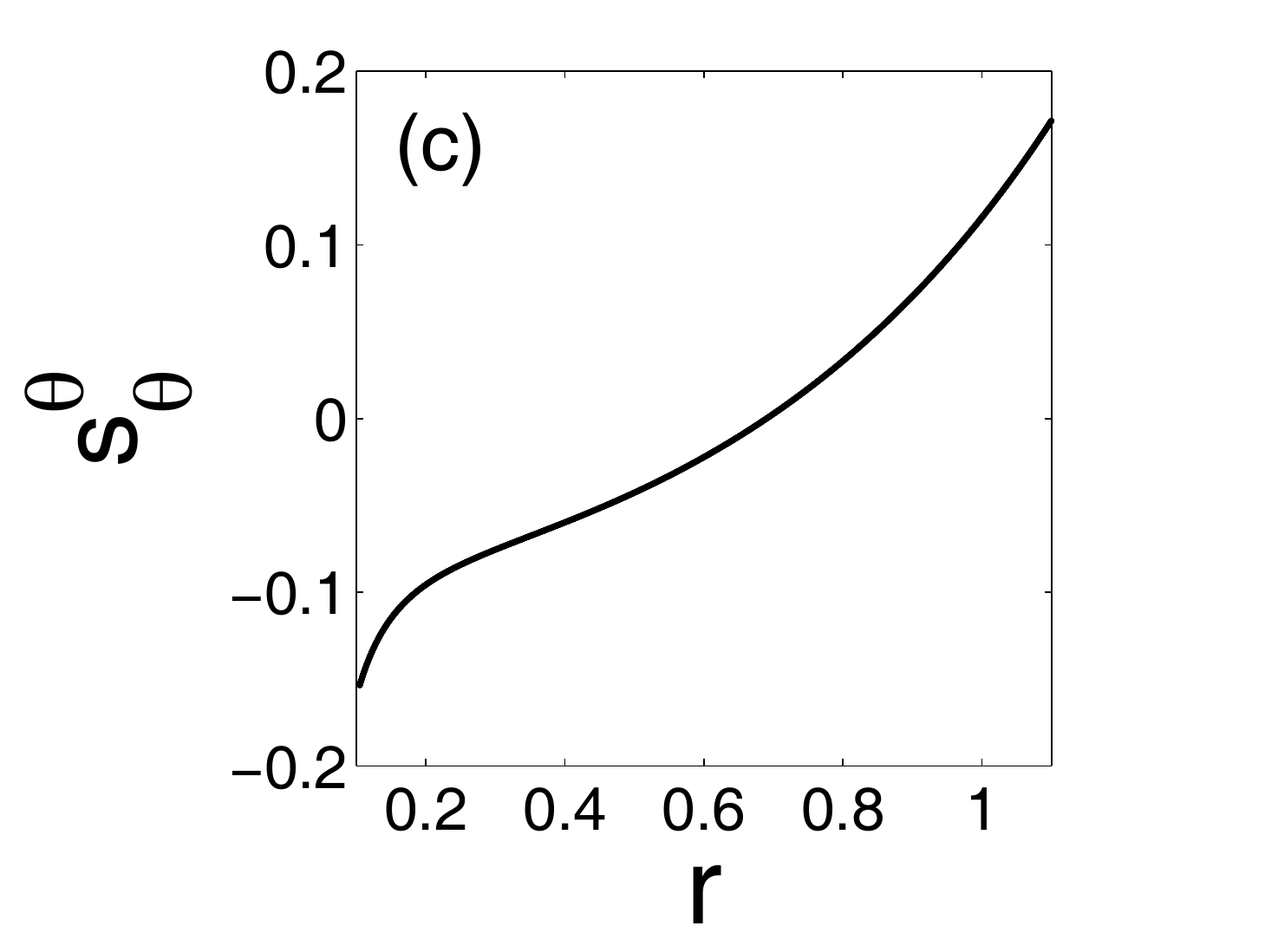}
\end{center}
\caption{(a) Plane-stress solution $\phi(r)$ for the elliptic metric \eqref{eq:elliptic} with Gaussian curvature $K=1$. (b)-(c) The corresponding principal stresses  $\stress^r_r(r)$ and $\stress^\theta_\theta(r)$.}
\label{fig:PS_Ell}
\end{figure}

\begin{figure}
\begin{center}
\includegraphics[height=\figsize in]{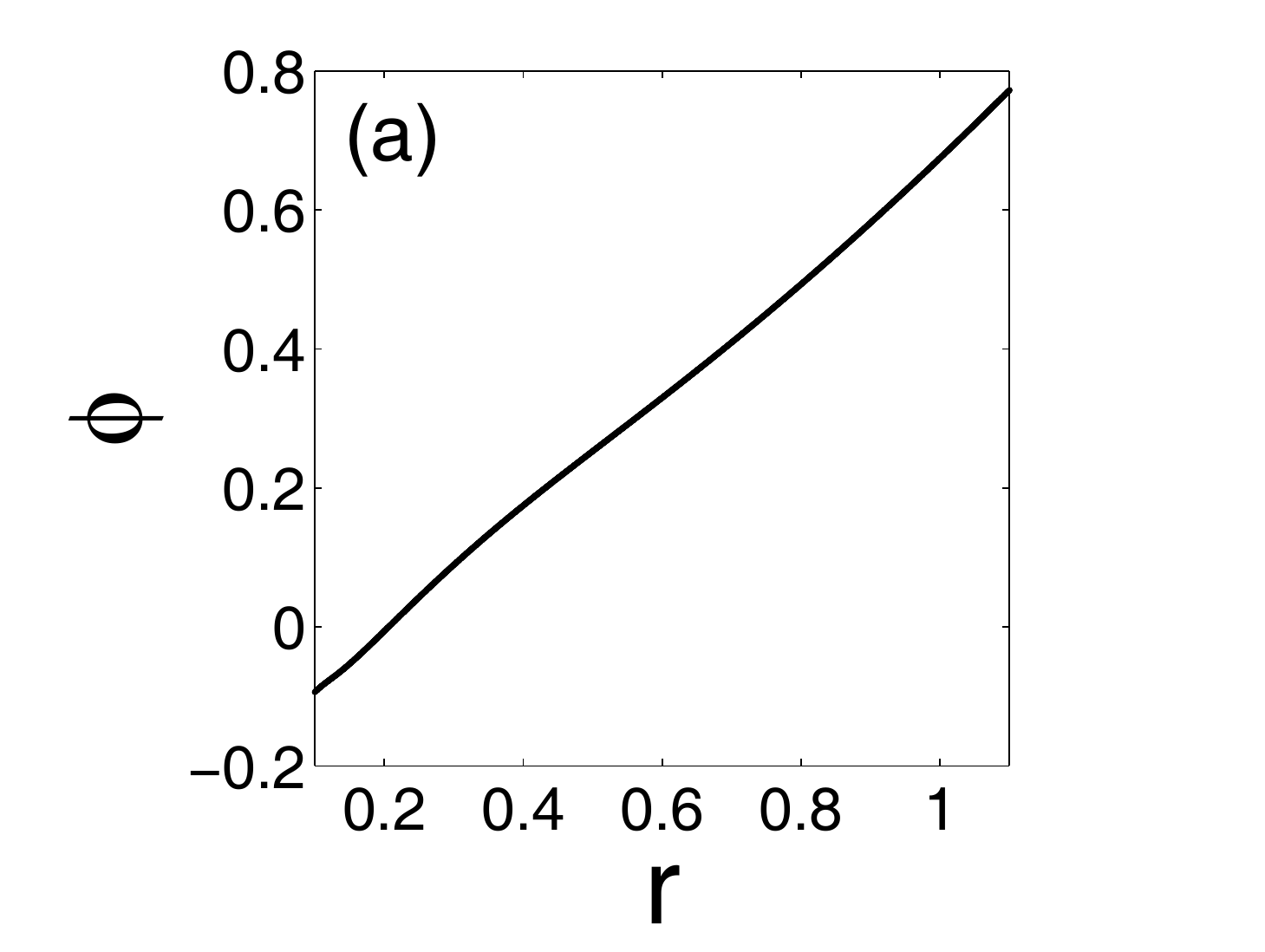}
\includegraphics[height=\figsize in]{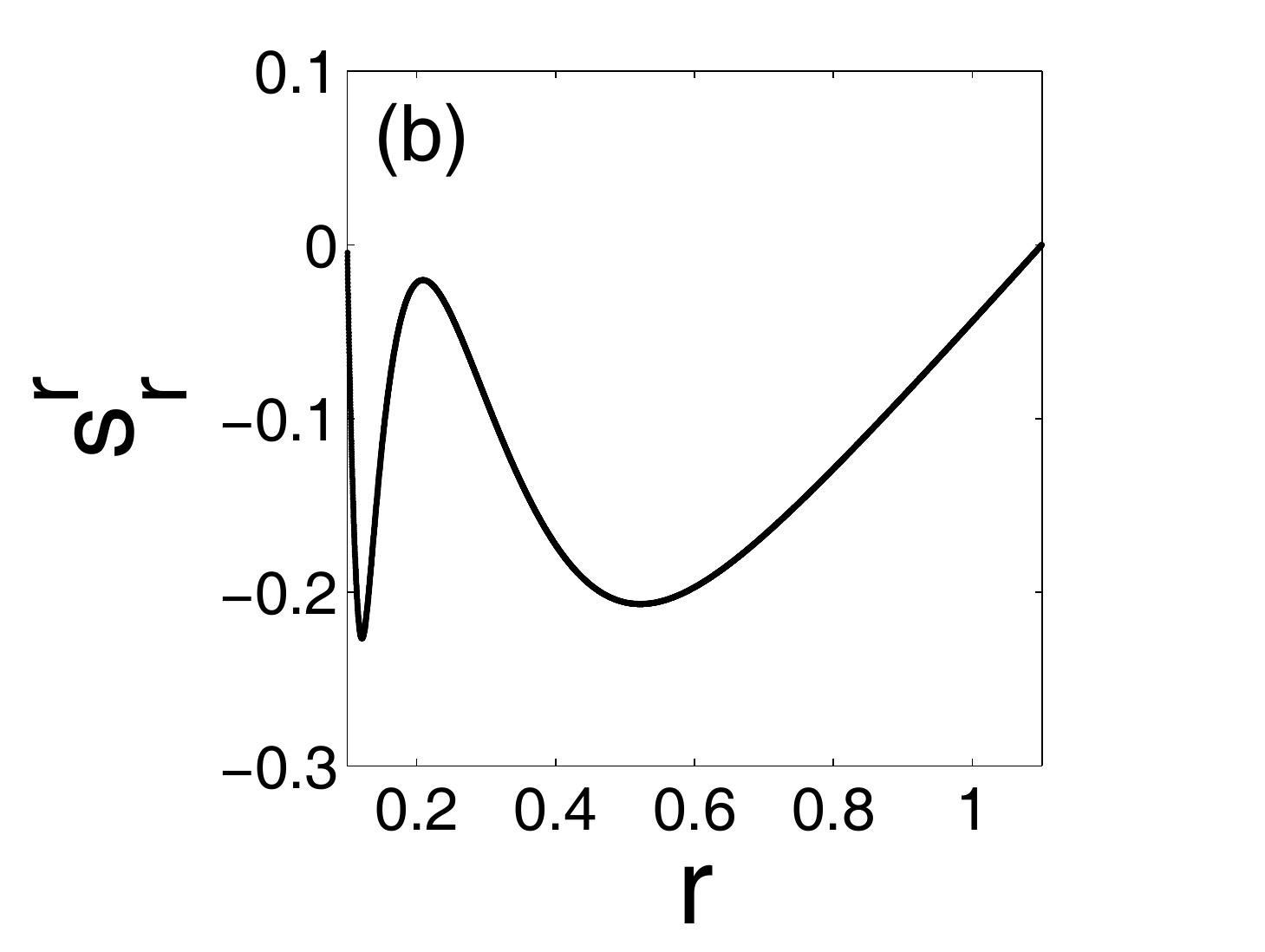}
\includegraphics[height=\figsize in]{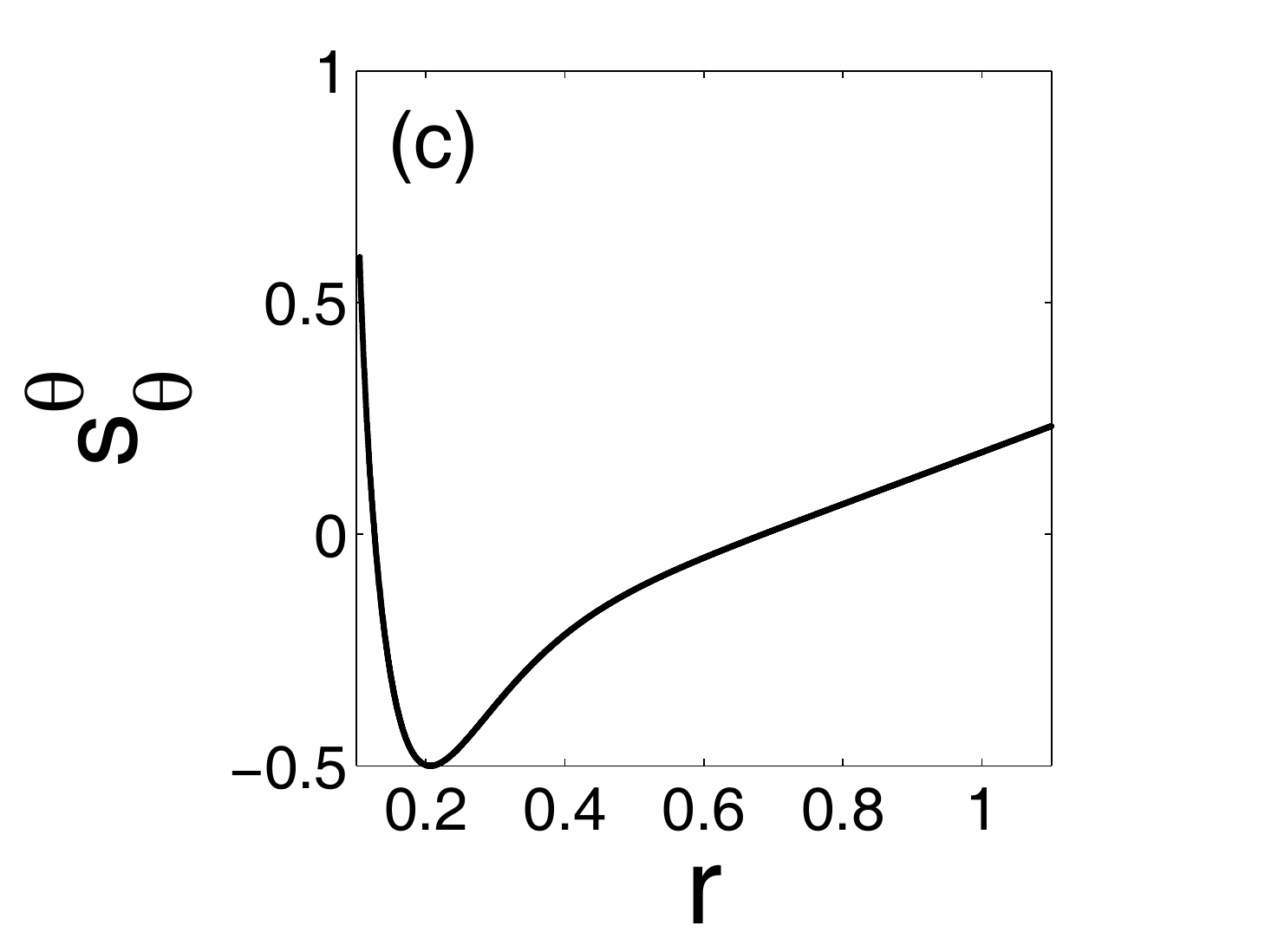}
\end{center}
\caption{(a) Plane-stress solution $\phi(r)$ for the flat metric \eqref{eq:flat} with $\alpha=0.58$. (b)-(c) The corresponding principal stresses $\stress^r_r(r)$ and $\stress^\theta_\theta(r)$.}
\label{fig:PS_Flat}
\end{figure}

\begin{figure}
\begin{center}
\includegraphics[height=\figsize in]{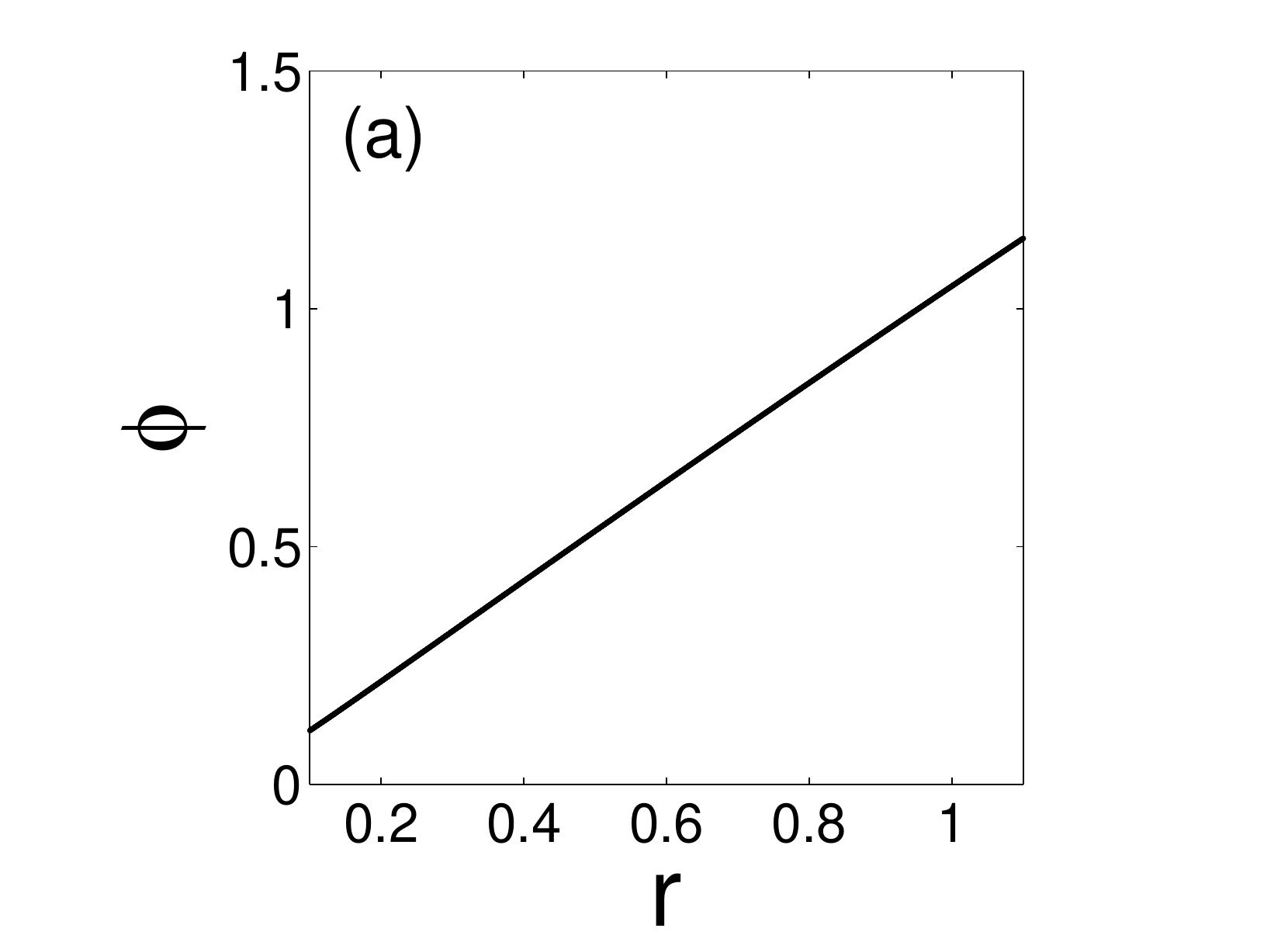}
\includegraphics[height=\figsize in]{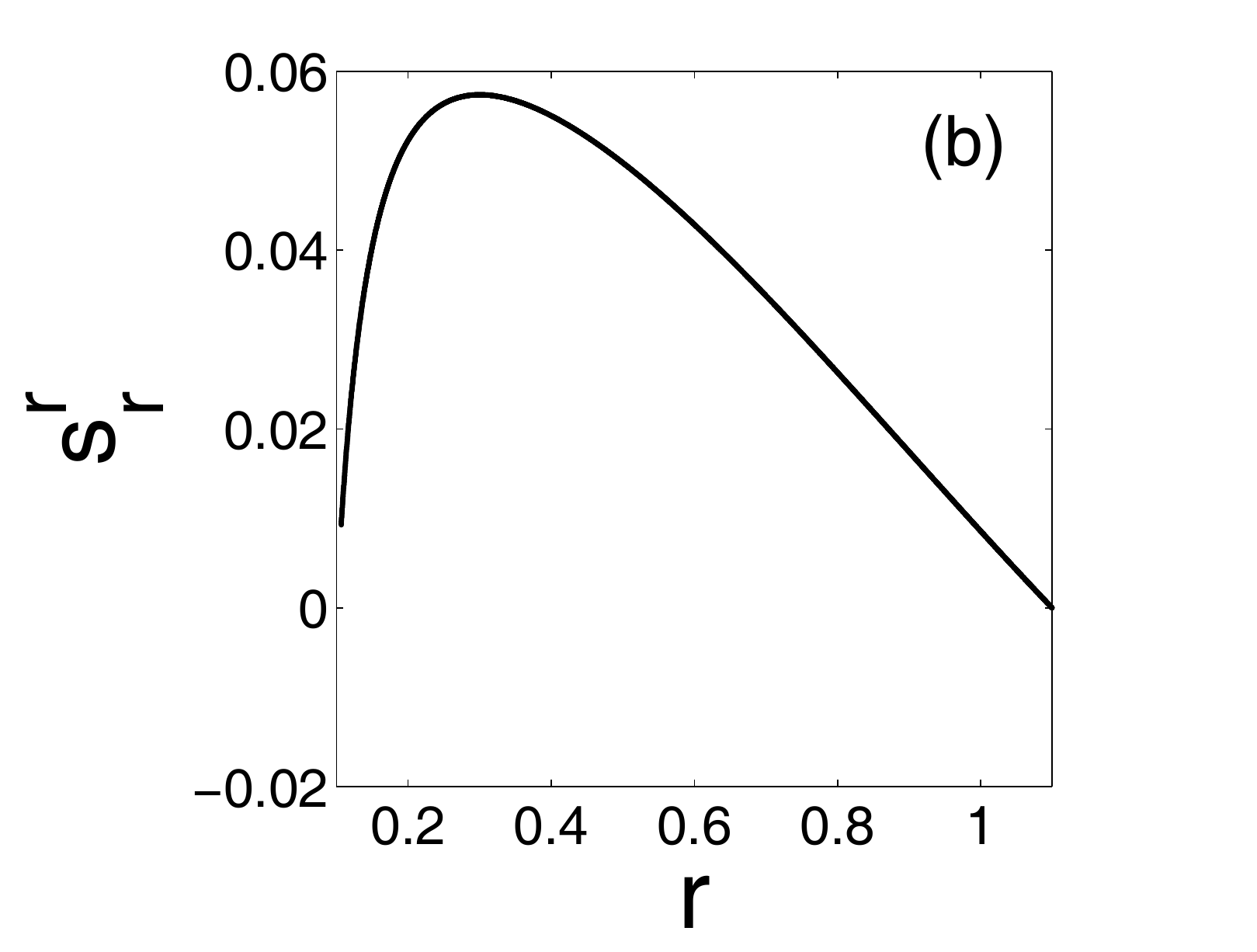}
\includegraphics[height=\figsize in]{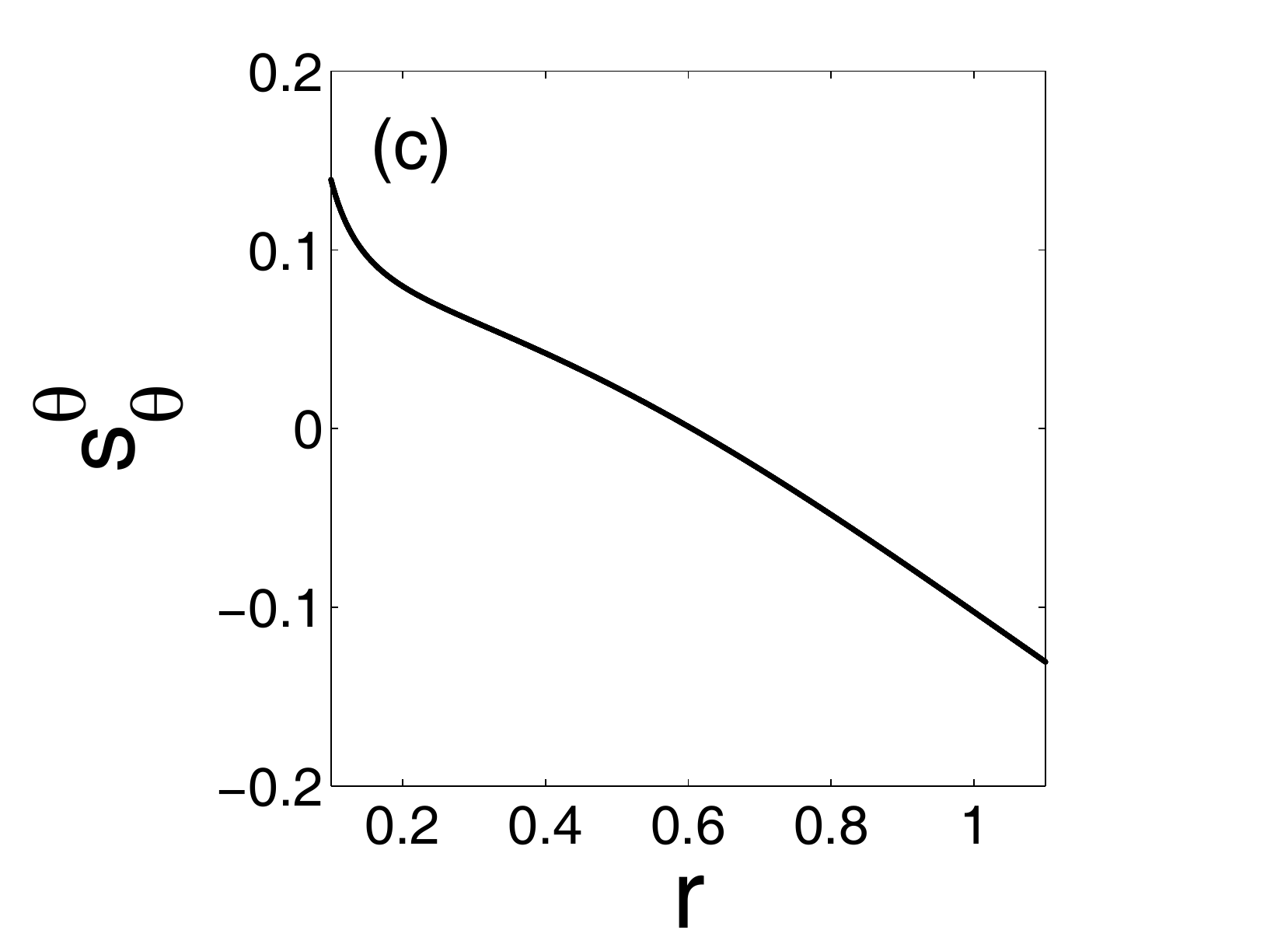}
\end{center}
\caption{(a) Plane-stress solution $\phi(r)$ for the hyperbolic metric \eqref{eq:hyperbolic} with Gaussian curvature $K=-1$. (b)-(c) The corresponding principal stresses $\stress^r_r(r)$ and $\stress^\theta_\theta(r)$.}
\label{fig:PS_Hyp}
\end{figure}

The plane-stress solutions are shown in Figures~\ref{fig:PS_Ell}--\ref{fig:PS_Hyp} for the domain $0.1\le r\le 1.1$. The solutions were obtained by a simple shooting procedure, with a fourth-order adaptive ODE solver.
For each metric we plot the solution $\phi(r)$ along with the spatial profile of the stress components $s_r^r(r)$ and $s^\theta_\theta(r)$ given by \eqref{eq:PSstress} up to the lowering of one index (the reason for displaying stress components with mixed upper and lower index is that only then all the components have the same units, hence can be compared). In the three cases, the solution $\phi(r)$ is close to linear. Significant differences are however observed in the stress components.

\begin{figure}
\begin{center}
\includegraphics[height=1.7in]{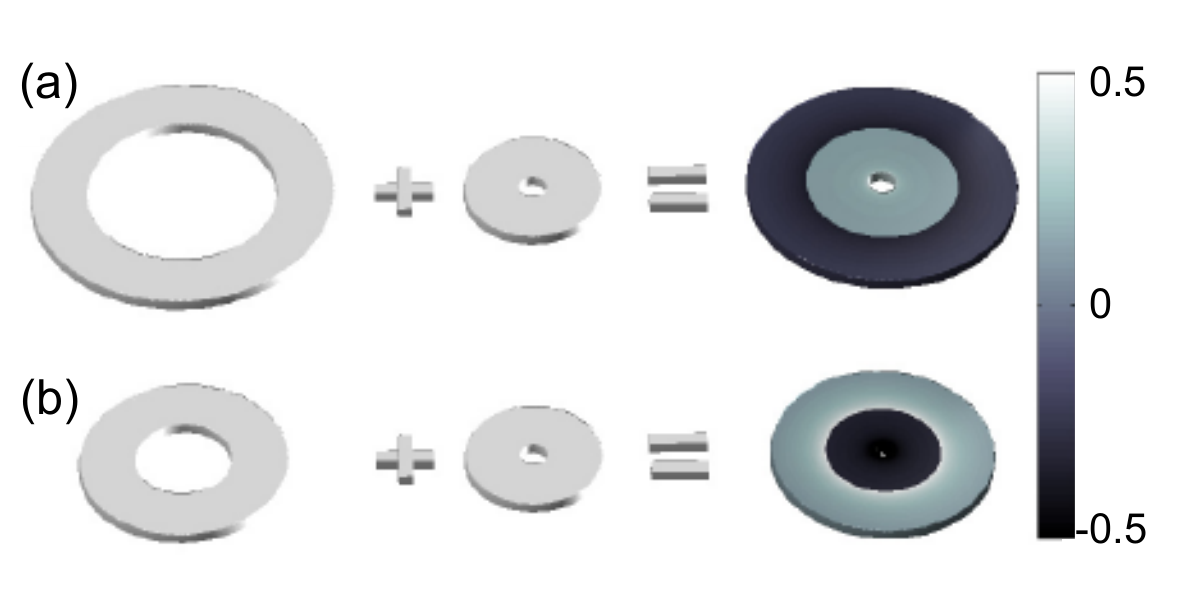}
\end{center}
\caption{
Cartoons of hyperbolic (a) and elliptic (b) plates.
In both cases, two punctured discs are ``glued" one inside the other and forced to remain planar. 
In the hyperbolic case, the inner perimeter of the outer disc is too long, compared with the outer perimeter of the inner disc. In such a case the inner disc is stretched azimuthally, while the outer disc is compressed azimuthally. In the elliptic case, the inner perimeter of the outer disc is too short, hence the inner disc is compressed azimuthally, while the outer disc is stretched azimuthally. The color bar on the right represents the azimuthal strain at equilibrium (computed numerically).  A similar cartoon was first presented in \cite{GB05}.
}
\label{fig:cartoon}
\end{figure}

For the elliptic metric (\figref{fig:PS_Ell}) the body is in a state of compression along the $r$ direction ($s^r_r<0$), whereas it is compressed in the $\theta$ direction near the inner radius and stretched along the $\theta$ direction near the outer radius. 
For the flat metric (\figref{fig:PS_Flat}) the situation is similar with compression everywhere along the radial direction, while  the angular stress switches from extension in the vicinity of the inner boundary to compression at larger radii, and again extension in the vicinity of the outer boundary. Finally, for the hyperbolic metric (\figref{fig:PS_Hyp}) the radial stress is everywhere positive (i.e., in a state of extension), whereas the angular stress is in a state of extension very close to the inner radius and in a state of compression at large enough distances form the center.
In  \figref{fig:cartoon} we show two toy models generating hyperbolic and elliptic geometries, which elucidate the behavior of the azimuthal (hoop) stress.

\subsection{Stability analysis and buckling threshold}

Let $\f(x^1,x^2)$ be the plane-stress configuration. Any small enough perturbation  can be decomposed into a sum of in-plane and out-of-plane displacements,
\[
\delta\f = v^\g \,\partial_\g\f + w\,\N,
\]
where $\N$ is the unit vector normal to the surface and $\partial_\g\f$ is the covariant derivative defined in \eqref{eq:cov_der}. Given such a perturbation we calculate in Appendix~\ref{sec:app} the variation in the elastic energy,
\[
\d E = \int_\calS \brk{\d w_S  + \thickness^2 \,\d w_B} \,\SurfaceElement,
\]
where the variation  in stretching content density, $\d w_S$, is given by \eqref{eq:dwS_full} and the variation in bending content density, $\d w_B$,  is given by \eqref{eq:dwB_full}; for flat surfaces these expression simplify considerably as $h_{\a\b}=0$.
Note that the plane-stress solution enters in the energy variation both through the stress $\stress^{\a\b}$ and through the metric parameters, $g_{\a\b}$ and $\Chr\g\a\b$.

The plane-stress solution is locally stable if the energy variation is positive for every choice of sufficiently small (non-trivial) perturbation.  As is well-known, local stability can be determined by considering only the leading order terms (in powers of $v,w$) in the energy variation.
The defining property of the plane-stress solution is that the terms that are linear in the in-plane perturbation $v^\g$ (the integral of $\d w_S^{(1,0)}$ in \eqref{eq:dwS_full}) vanish for every choice of $v^\g$. Thus, to leading order, the energy variation decomposes into a sum of terms that are quadratic in $v$ and terms that are quadratic in $w$,
\[
\begin{split}
\d  E &= \d E_S^{(2,0)}(v) +   \Brk{\d E_S^{(0,2)}(w)
+  \thickness^2\, \d E_B^{(0,2)}(w)}  + \O(v^3, v^2 w, v w^2, w^3),
\end{split}
\]
where
\beq
\begin{aligned}
\d E_S^{(2,0)}(v) &=
\half\int_\calS  \Big\{ \stress^{\a\b}  g_{\g\e} (\nabla_\a v^\g)(\nabla_\b v^\e)  \\
&\quad +
 \calA^{\a\b\g\d} g_{\b\eta} g_{\d\e}  (\nabla_\a v^\eta) (\nabla_\g v^\e) \Big\} \,\SurfaceElement \\
\d E_S^{(0,2)}(w) &= \int_\calS  \half \stress^{\a\b} (\nabla_\a w) (\nabla_\b w) \,\SurfaceElement \\
\d E_B^{(0,2)}(w) &= \int_\calS
\frac{1}{24} \calA^{\a\b\g\d}  (\nabla_\b\nabla_\a w)  (\nabla_\d\nabla_\g w) \,\SurfaceElement
\end{aligned}
\label{eq:dE_buck}
\eeq
(the subscripts $(i,j)$ refer to the power of $v$ and $w$).

Since, to leading order, the energy variation is decomposed into a sum of a $v$-dependent term and a $w$-dependent term, the minimization can be performed on each component separately. By assumption, the plane-stress solution is the energy minimizer with respect to in-plane perturbations, thus minimum energy variation is obtained for $v^\g=0$.

It remains to consider the energy variation due to out-of-plane perturbations.
The bending term $\d E_B^{(0,2)}(w)$ is always positive due to the positive-definiteness of the tensor $\calA^{\a\b\g\d}$\,\footnote{
For any tensor $a_{\a\b}$ we have
\[
\calA^{\a\b\g\d} a_{\a\b} a_{\g\d} = \frac{\nu}{1-\nu} a_\a^\a a_\b^\b + a_\a^\b a_\b^\a,
\]
which is positive for all $0\le\nu<1$ when $a_{\a\b}$ is symmetric.
}.
Whether the stretching term $\d E_S^{(0,2)}(w)$ is sign-definite depends on the plane-stress solution. In fact, if the stress tensor is not \emph{everywhere} positive-definite, then there exists a perturbation $w$ for which $\d E_S^{(0,2)}(w)$ is negative, and by taking the plate thickness $\thickness$ sufficiently small, the total energy variation can be made negative. We have thus recovered the following general result:

\begin{quote}
\emph{Given a reference metric, the plane-stress solution is linearly stable against buckling, independently of the plate thickness, only if the stress is everywhere positive-definite. In other words, an infinitely thin plate cannot sustain compression without buckling. }
\end{quote}

We will now show that the existence of a buckling threshold is always guaranteed, unless the plane-stress solution is trivial, i.e., $\stress^{\a\b}=0$ (which in turn occurs only if the reference metric is flat). We start by noting that the plane-stress equations \eqref{eq:plane_stress} can be rewritten as
\[
\nabla_\b\brk{\frac{\sqrt{|\go|}}{\sqrt{|g|}} \stress^{\a\b}} = 0,
\]
where $\nabla_\b$ is the covariant derivative defined in the appendix.
Let now $\chi(x^1,x^2)$ be a scalar field satisfying $\nabla_\b\nabla_\a\chi=0$ and consider the integral
\[
I = \int_\calS \stress^{\a\b} (\nabla_\a\chi)  (\nabla_\b\chi)\,\SurfaceElement.
\]
The surface element $\SurfaceElement$ is defined in terms of the reference metric. Writing $\SurfaceElement = (\sqrt{|\go|}/\sqrt{|g|})\sqrt{|g|} \,dx^1 dx^2$, we may now integrate by parts (the covariant derivative satisfies the usual rules of integration by parts provided that the surface element is consistent with the Christoffel symbols), using the boundary conditions $\stress^{\a\b} n_\b=0$,
\[
I = \int_\calS \nabla_\b\brk{\frac{\sqrt{|\go|}}{\sqrt{|g|}} \stress^{\a\b} (\nabla_\a\chi)}  \chi \sqrt{|g|} \,dx^1 dx^2.
\]
Since the covariant derivative satisfies the Leibniz rule for the derivative of products, it follows from the plane-stress equations and the definition of $\chi$ that $I=0$.

Thus, if there exists a scalar function $\chi$ that has a non-zero (covariant) gradient and satisfies $\nabla_\a\nabla_\b\chi=0$, then the fact that $I=0$ implies that $s^{\a\b}$ is not everywhere positive-definite. A simple way to show that such a function does exist is to endow the planar equilibrium state with a Cartesian set of coordinates. Then the covariant derivative reduces into a simple partial derivative, and the function, say, $\chi(x^1,x^2) = x^1$ has the desired property.

We may summarize as follows:

\begin{quote}
\emph{A sufficiently thin unconstrained non-Euclidean plate will always buckle unless the plane-stress solution is trivial, i.e., $\stress^{\a\b}=0$.}
\end{quote}

Equation \eqref{eq:dE_buck} provides a characterization of the critical thickness $\thickness=\tc$ at which buckling first occurs. At criticality, $\thickness=\tc$, there exists a non-trivial (i.e., non-uniform) perturbation which to leading order is marginally unstable,  i.e.,
\[
\inf_{w\ne\text{const}}  \int_{\calS} \Big\{\half \stress^{\a\b} (\nabla_\a w) (\nabla_\b w) +
\frac{\tc^2}{24} \calA^{\a\b\g\d} (\nabla_\b\nabla_\a w)  (\nabla_\d\nabla_\g w) \Big\} \,\SurfaceElement
 = 0,
\]
which implies that
\beq
\tc^2 = \sup_{w\ne\text{const}}
\frac{-12 \int_{\calS} \stress^{\a\b} (\nabla_\a w) (\nabla_\b w) \,\SurfaceElement}{
\int_{\calS} \calA^{\a\b\g\d} (\nabla_\b\nabla_\a w)  (\nabla_\d\nabla_\g w) \,\SurfaceElement}.
\label{eq:hc}
\eeq
By the above analysis
this supremum is guaranteed to be non-negative, and zero if and only if  $\stress^{\a\b}=0$.

\parag{Comments:}
\begin{enumerate}
\item Equation \eqref{eq:hc} provides a mean for generating lower bounds for the buckling threshold by choosing appropriate trial functions, $w$.

\item The energy variation \eqref{eq:dE_buck} is a quadratic functional of $w$ of the form,
\beq
\d E_S^{(0,2)}(w) + \thickness^2\,\d E_B^{(0,2)}(w) = (w,\calH w),
\label{eq:calH}
\eeq
where $(\cdot,\cdot)$ is the standard inner-product on $\calS$ and $\calH$ is a self-adjoint second-order differential operator. Above the buckling threshold $\calH$ is positive-definite. The buckling threshold $\tc$ corresponds to the largest $\thickness$ for which $\calH$ has a zero eigenvalue. From a numerical point of view, the latter characterization is the easier way for computing the buckling threshold.

\end{enumerate}

\parag{Examples}
We turn back to the punctured discs considered in the previous subsection. Note that in all three cases there exist negative stress components, hence a buckling transition is guaranteed to occur at some finite thickness.

We denote by $\phi(r)$ the solution to the plane-stress equation \eqref{eq:plane-stress-example}. For an out-of-plane perturbation $w(r,\theta)$, substituting  \eqref{eq:Chr_example}, we get
\[
(\nabla_\b\nabla_\a w) = (\partial_\a\partial_\b w) - \Chr\eta\a\b (\partial_\eta w) =
\mymat{\phi' (w'/\phi')' & \phi (\dot{w}/\phi)' \\
\phi (\dot{w}/\phi)' & \ddot{w} + \phi (w'/\phi')},
\]
where we denote by primes derivatives with respect to $r$ and by dots derivatives with respect to $\theta$.
Thus,
\begin{widetext}
\beq
\begin{aligned}
\d E_S^{(0,2)}(w) &= \half \int_0^{2\pi} \int_{\rmin}^{\rmax}  \brk{\stress^{rr}  (w')^2 + \stress^{\theta\theta} \frac{\dot{w}^2}{\Phi^2}}\Phi\,dr d\theta \\
\d E_B^{(0,2)}(w) &= \frac{1}{24} \int_0^{2\pi} \int_{\rmin}^{\rmax}  \BRK{
(\phi')^2 \Brk{\brk{\frac{w'}{\phi'}}'}^2 + 2  \frac{\phi^2}{\Phi^2} \Brk{\brk{\frac{\dot{w}}{\phi}}'}^2 +
\frac{\phi^2}{\Phi^4} \brk{\frac{\ddot{w}}{\phi} +\frac{w'}{\phi'}}^2}\Phi\,dr d\theta,
\end{aligned}
\label{eq:dE20}
\eeq
\end{widetext}
with $\stress^{rr}$ and $\stress^{\theta\theta}$ given by  \eqref{eq:PSstress}. Due to the periodicity in $\theta$
it is natural to expand the perturbation in Fourier series,
\[
w(r,\theta) = a_0(r) + \sqrt{2} \sum_{n=1}^\infty a_n(r) \,\cos n\theta + \sqrt{2} \sum_{n=1}^\infty b_n(r) \,\sin n\theta.
\]
Because \eqref{eq:dE20} is quadratic in $w$, both terms reduce into a sum over Fourier components,
\[
\begin{aligned}
\d E_S^{(0,2)}(w) &=  \sum_{n=0}^\infty [\d E_S^n(a_n) + \d E_S^n(b_n)]  \\
\d E_B^{(0,2)}(w) &=  \sum_{n=0}^\infty [\d E_B^n(a_n)+ \d E_B^n(b_n)] ,
\end{aligned}
\]
where we define for every function $z =z(r)$,
\begin{widetext}
\[
\begin{aligned}
\d E_S^n(z) &= \half  \int_{\rmin}^{\rmax}  \BRK{\stress^{rr} (z')^2 + \stress^{\theta\theta} (n z)^2}\Phi\,dr  \\
\d E_B^n(z) &= \frac{1}{24}  \int_{\rmin}^{\rmax}  \BRK{
(\phi')^2 \Brk{\brk{\frac{z'}{\phi'}}'}^2 + 2  \frac{\phi^2}{\Phi^2} \Brk{\brk{\frac{n z}{\phi}}'}^2 +
\frac{\phi^2}{\Phi^4} \brk{\frac{n^2 z}{\phi} + \frac{z'}{\phi'}}^2}\Phi\,dr .
\end{aligned}
\]
\end{widetext}
The buckling threshold \eqref{eq:hc} is given by
\[
\tc^2 = \sup_{\{a_n,b_n\}} \frac{-\sum_{n=0}^\infty [\d E_S^n(a_n)+\d E_S^n(b_n)]}{
\sum_{n=0}^\infty [\d E_B^n(a_n)+\d E_B^n(b_n)]}.
\]

\parag{Corollary 1}
Let
\[
(\thickness_n^*)^2 = \sup_z \frac{-\d E_S^n(z)}{\d E_B^n(z)}.
\]
Then it is clearly the case that $\thickness_n^* \le \tc$ for every $n$. On the other hand, $\tc \le \max_n \thickness_n^*$, which together implies that
\[
\tc^2 = \max_n \sup_z \frac{-\d E_S^n(z)}{\d E_B^n(z)}.
\]
Thus, unless the buckling transition is degenerate, then the marginally stable perturbation at the bifurcation point involves a single Fourier mode.

\parag{Corollary 2}
A buckling transition occurs if either $\stress^{rr}$ or $\stress^{\theta\theta}$ are somewhere negative. Suppose that $\stress^{rr}(r)>0$, i.e., the radial stress is everywhere extensional. It follows that $\d E_S^0(z)\ge 0$ for every $z$ (every axisymmetric perturbation, $n=0$, increases the stretching energy). If $\stress^{\theta\theta}$ is somewhere negative, then there exist non-axisymmetric perturbations that reduce the elastic energy. That is, the buckling transition breaks the axial symmetry.

\begin{figure}
\begin{center}
\includegraphics[height=\figsize in]{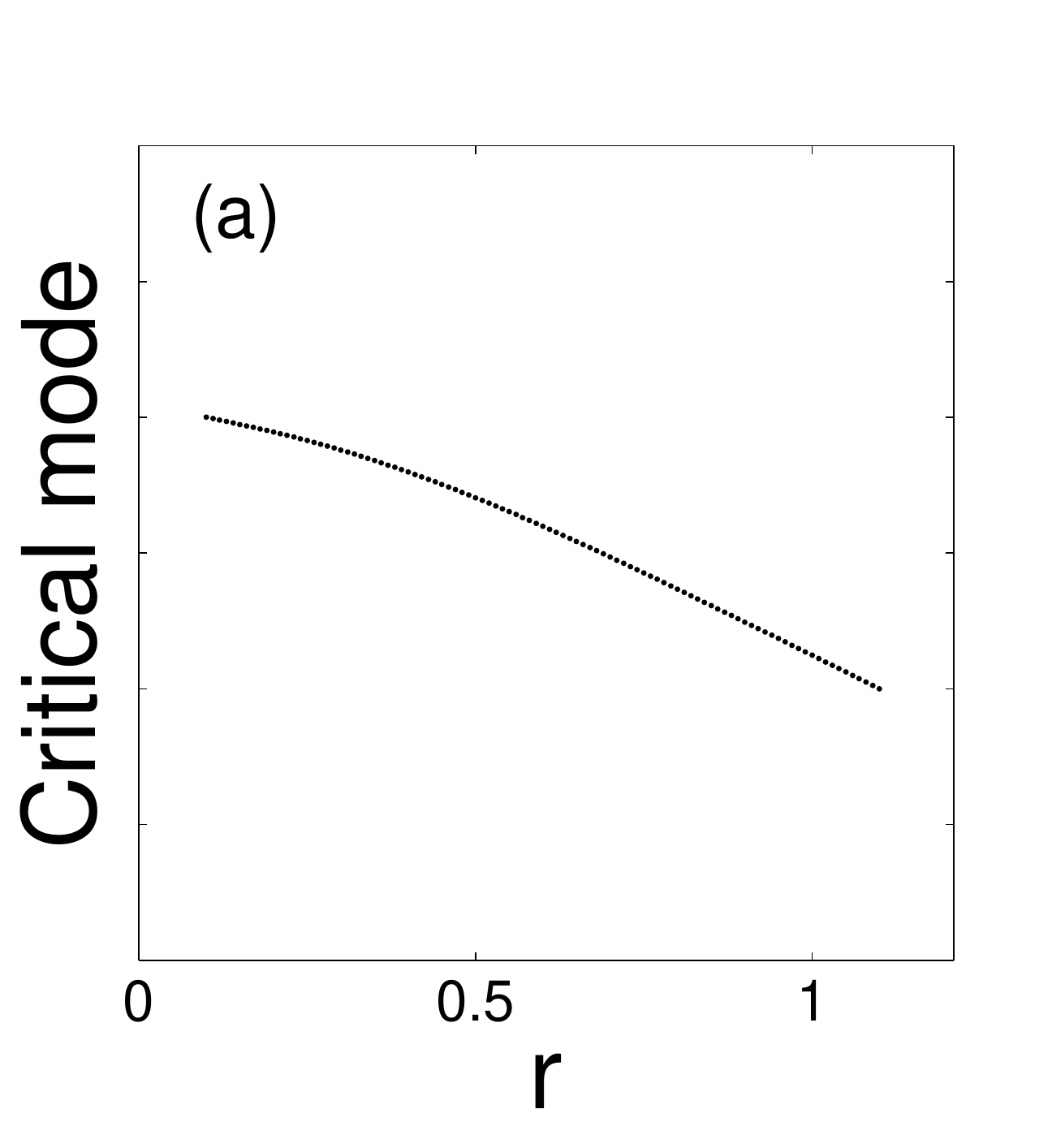}
\includegraphics[height= \figsize in]{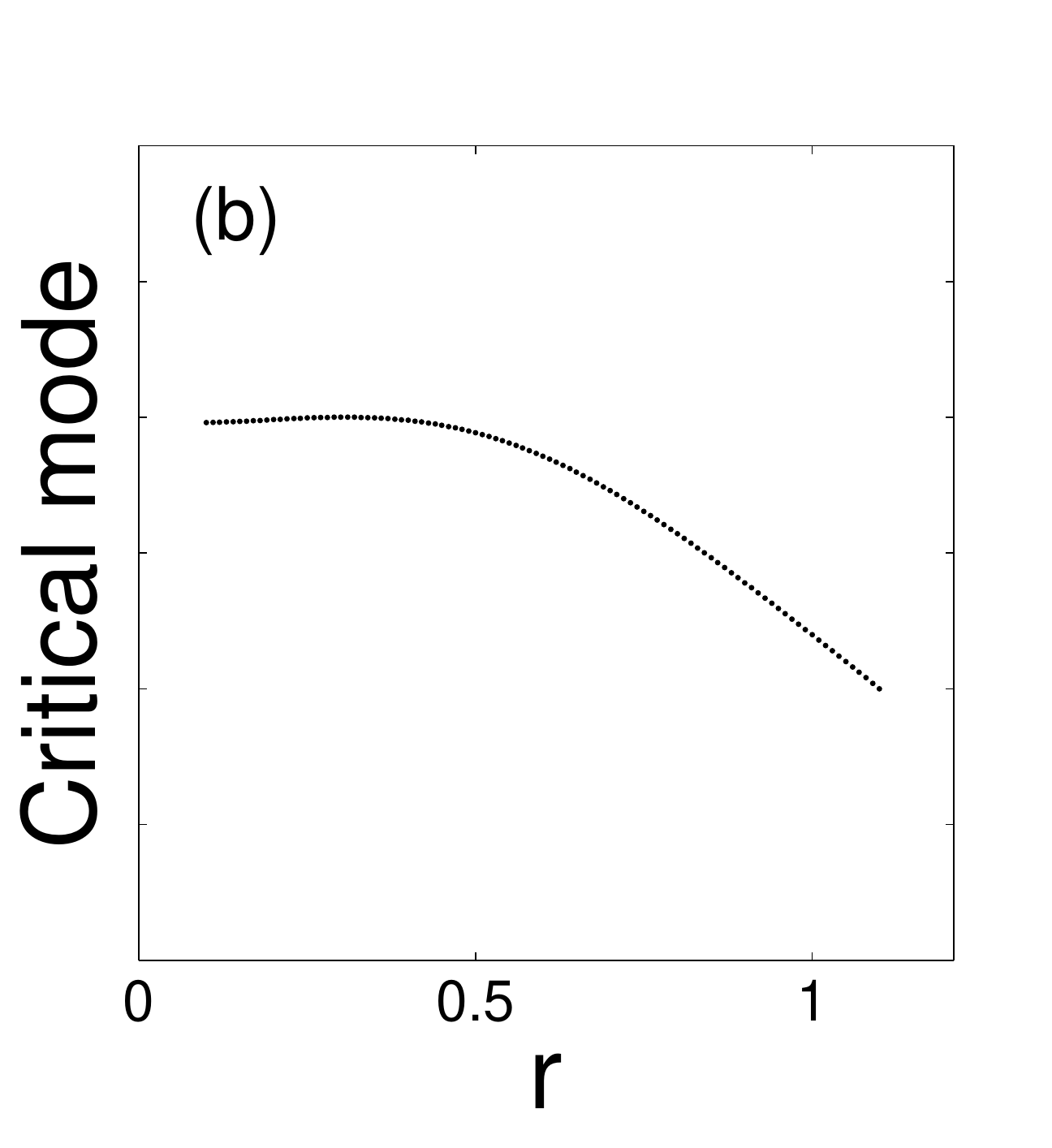}
\includegraphics[height= 1.75 in]{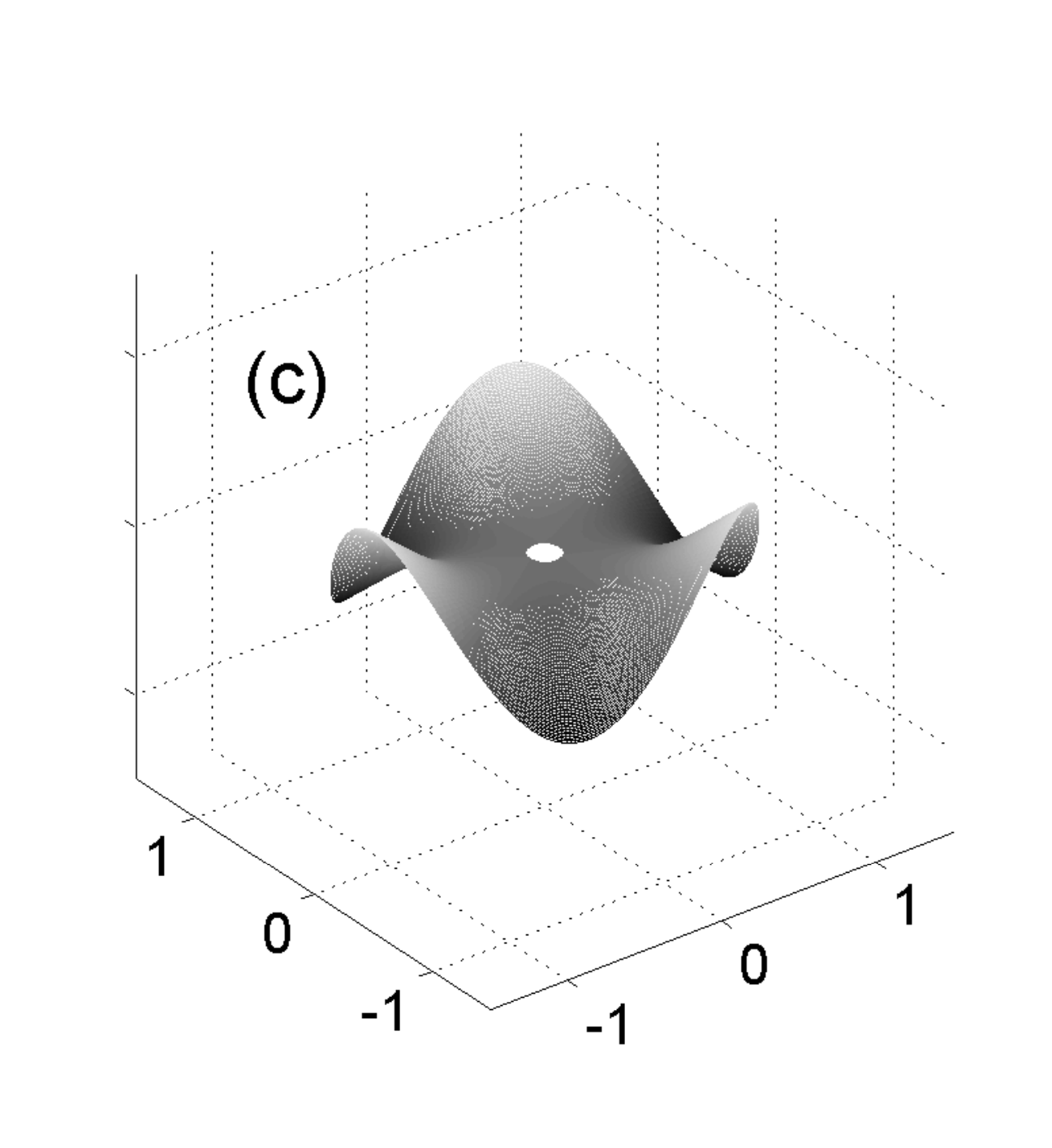}
\end{center}
\caption{Critical modes for the elliptic (a), flat (b) and hyperbolic (c) geometries. For both elliptic and flat geometries the critical mode is axisymmetric ($n=0$), hence we only display a cross section. In the hyperbolic case the first mode to destabilize is $n=3$. }
\label{fig:Modes}
\end{figure}

\parag{Elliptic geometry}
Consider first the elliptic geometry \eqref{eq:elliptic} for the same parameters as in \figref{fig:PS_Ell}. The buckling threshold occurs at $\tc=0.367$, and corresponds to an axisymmetric mode ($n=0$). The critical mode is shown in \figref{fig:Modes}$a$.  In \tabref{tab:elliptic} we show the buckling threshold $\tc$ versus the Gaussian curvature $K$. As expected, the buckling threshold is higher the more curved the surface is.

\begin{table}
\begin{center}
\begin{tabular}{|r|rrrrrrr|}
\hline
$K$ & 0.2 & 0.4 & 0.6 & 0.8 & 1.0 & 1.2 & 1.4 \\
\hline
$\tc$ & 0.164 & 0.233 & 0.285 & 0.329 & 0.367 & 0.401 & 0.431 \\
\hline
\end{tabular}
\end{center}
\caption{Buckling threshold $\tc$ versus the Gaussian curvature $K$ for the elliptic geometry \eqref{eq:elliptic}.}
\label{tab:elliptic}
\end{table}

\parag{Flat geometry}
Consider next the flat geometry \eqref{eq:flat} for the same parameters as in \figref{fig:PS_Flat}. The buckling threshold occurs at $\tc=0.387$, also for an axisymmetric mode. The critical mode is shown in \figref{fig:Modes}$b$.

\parag{Hyperbolic geometry}
Consider finally the hyperbolic geometry \eqref{eq:hyperbolic} for the same parameters as in \figref{fig:PS_Hyp}.
Since $\stress^{rr}>0$ it follows that the critical mode must break the polar symmetry. Indeed, the least stable mode, which changes stability at $\tc=0.1845$, has harmonic $n=3$. It is depicted in \figref{fig:Modes}$c$. Note how lower is the buckling threshold for the hyperbolic geometry. Finally, we show in \tabref{tab:hyperbolic} the buckling threshold $\tc$ versus the Gaussian curvature $K$.

\begin{table}
\begin{center}
\begin{tabular}{|r|rrrrrrrrrr|}
\hline
$-K$ & 0.2 & 0.4 & 0.6 & 0.8 & 1.0 & 1.2 & 1.4 & 1.6 & 1.8 & 2.0\\
\hline
$\tc$ & 0.0768 & 0.110 & 0.135 & 0.157 & 0.175 & 0.192 & 0.208 & 0.222 & 0.236 & 0.248 \\
\hline
\end{tabular}
\end{center}
\caption{Buckling threshold $\tc$ versus the Gaussian curvature $-K$ for the hyperbolic geometry \eqref{eq:hyperbolic}. In all cases the critical mode has harmonic $n=3$.}
\label{tab:hyperbolic}
\end{table}

\subsection{Buckling threshold versus crossover point}

Equation \eqref{eq:hc}  expresses the buckling threshold $\tc$ as a supremum over trial normal deflections. As such, it provides an easy way to generate lower bounds for the buckling threshold.  An approximation often used to estimate the buckling threshold is the so-called crossover point between the lowest-energy isometric immersion and the plane-stress solution. In this section we show that the crossover point can often yield a significant underestimate to the buckling transition.

The equilibrium configuration is the one that minimizes the energy functional \eqref{eq:Efunc}. Two upper bounds that correspond to extreme cases are the plane-stress solution, which involves zero bending energy, and the isometric immersion that minimizes the Willmore functional, which  involves zero stretching energy. If $g_{\a\b}^{\text{PS}}$ denotes the plane-stress metric then
\[
E  = \frac{1}{8} \int_\calS \calA^{\a\b\g\d}(g_{\a\b}^{\text{PS}} - \go_{\a\b})(g_{\g\d}^{\text{PS}} - \go_{\d\g})
\,\SurfaceElement \equiv E_{\text{PS}},
\]
whereas if $h_{\a\b}^{\text{WF}}$ is the second quadratic form that minimizes the Willmore functional (subject to the satisfaction of the Gauss-Mainardi-Codazzi equations with $g_{\a\b}=\go_{\a\b}$)  then the energy reduces into 
\[
E = \frac{\thickness^2}{24} \int_\calS \calA^{\a\b\g\d} h_{\a\b}^{\text{WF}}  h_{\g\d}^{\text{WF}} \,\SurfaceElement
\equiv \thickness^2\,E_{\text{WF}} .
\]
Clearly, if the Willmore energy is lower than the plane-stress energy then the plane-stress solution is unstable. This provides a lower bound for the buckling threshold, known as the crossover point,
\[
\tc \ge \sqrt{\frac{E_{\text{PS}}}{E_{\text{WF}}}} \equiv \tco.
\]
Below this thickness, we expect the solution to approach an isometric immersion of minimum energy, thus the energy should be close to $E_{\text{WF}}$.
Obviously, in order to evaluate the crossover point one needs to know the minimizer of the Willmore functional, which may be highly non-trivial (it requires in particular the solution of an isometric immersion problem).

\parag{Examples}
Consider the elliptic and flat geometries \eqref{eq:elliptic} and \eqref{eq:flat}, for which the isometric immersion that minimizes the Willmore functional is explicitly known.  It is a surface of revolution,
\beq
\f(r,\theta) = (\phi(r)\,\cos\theta, \phi(r)\,\sin\theta,\psi(r)).
\label{eq:revolution}
\eeq
The corresponding metric is
\[
g_{\a\b} = \mymat{(\phi')^2 + (\psi')^2 & 0 \\ 0 & \phi^2},
\]
hence the isometric immersion satisfies
\[
\phi = \Phi \Textand (\phi')^2 + (\psi')^2=1.
\]
(For the hyperbolic metric \eqref{eq:hyperbolic} $\Phi'(r)>1$ hence there is no axisymmetric isometric immersion.)

For the elliptic geometry \eqref{eq:elliptic} with the same parameters as above we find
\[
E_{\text{PS}} = 0.0163 \Textand
E_{\text{WF}} = 0.3609,
\]
from which we get $\tco=0.2125$, which is lower than $\tc=0.367$ by about 40\%. In contrast, we obtain for the flat metric \eqref{eq:flat}
\[
E_{\text{PS}} = 0.0580 \Textand
E_{\text{WF}} = 75.64,
\]
from which we get $\tco=0.0277$, which is lower than $\tc=0.387$ by more than an order of magnitude. This demonstrates that in certain cases the crossover point may provide a very poor estimate of the buckling threshold. The reason why the discrepancy between $\tco$ and $\tc$ may be large is that the buckling transition is a property intrinsic to  the plane-stress solution, not to isometric immersions.

\subsection{Bifurcation analysis}

In this section we analyze the nature of the buckling transition. At $\thickness=\tc$ the plane-stress solution is marginally stable. In particular, there exists a non-trivial perturbation that does not change the elastic energy up to terms that are quadratic in $v,w$. Specifically,
\[
\d E_S^{(2,0)}(v)=0 \qquad\text{ if and only if $v=0$},
\]
and there exists a $\tw\ne0$ such that
\[
\d E_S^{(0,2)}(\tw) + \tc^2 \,\d E_B^{(0,2)}(\tw)
\]
changes sign at $\thickness=\tc$. In fact, as $\d E_B^{(0,2)}$ can be identified as an inner-product (cf. \eqref{eq:calH}), it follows that for every $\hat{w}$,
\beq
(\hat{w},\calH \tw) = 0.
\label{eq:tww}
\eeq
Since $\tw$ is determined up to both additive and multiplicative constants, we will define $\tw$ to have zero mean and be normalized,  $\|\tw\|_2=1$, where $\|\cdot\|_2$ is the $L^2$ norm.

For plate thickness below $\tc$ the flat configuration is linearly unstable. (Note however that the plane-stress solution is a critical point of the energy functional for all values of $\thickness$; it only ceases to be a local minimum at $\tc$).
The loss of stability of the flat solution is due to a bifurcation. A branch of stationary solutions with non-zero bending content merges with the plane-stress solution at $\thickness=\tc$. The bifurcation is called super-critical (forward) if the branch of buckled solutions exists for $\thickness\le \tc$, in which case, as predicted by bifurcation theory \cite{IJ97}, the buckled solutions near $\tc$ are linearly stable. For a super-critical bifurcation
the transition from the plane-stress solution to the buckled solution, as $t$ decreases below $\tc$,  is continuous.
The bifurcation is called sub-critical (backward)  if the branch of buckled solutions exists for $\thickness\ge \tc$. In this case, the buckled solutions near $\tc$ are unstable (this branch of solutions becomes stable after it turns back). A transition to linearly stable solutions occurs discontinuously at $\thickness=\tc$. In particular, discontinuous bifurcations exhibit hysteresis.

To analyze the bifurcation we need to study the behavior of the energy functional in the vicinity of the bifurcation threshold. Since the plane-stress solution is marginally stable at $\tc$, terms that are of higher order in $v,w$ must be taken into account.

Let $g_{\a\b}$ and $s_{\a\b}$ be the plane-stress metric and stress, and set $\thickness^2 =\tc^2-\e$ with $\e>0$ a small parameter. That is, we consider plate thicknesses just below the buckling threshold.
By the above discussion, the bifurcation is super-critical if for small $\e$ the energy functional has a local minimum for a non-trivial perturbation whose magnitude vanishes as $\e\downarrow0$.
If the bifurcation is sub-critical, then the stable solution for $\e>0$ does not converge to the plane-stress solution as $\e \downarrow0 $.
Our working hypothesis is that the bifurcation is super-critical. The analysis will prove us wrong if this is not the case.

Set once again $\d\f = v^\g\,\df\g + w\N$. Substituting the variations \eqref{eq:dwS_full}, \eqref{eq:dwB_full} in stretching and bending content densities, the variation in total energy takes the form
\[
\begin{split}
\d E &= \d E_S^{(2,0)}(v) + \d E_S^{(0,2)}(w) + \d E_S^{(1,2)}(v,w) + \d E_S^{(0,4)}(w) \\
&+ (\tc^2-\e)\BRK{\d E_B^{(0,2)}(w) + \d E_B^{(1,2)}(v,w) + \d E_B^{(0,4)}(w)} \\
&+ \O(v^3,v^2 w, v w^3, w^5),
\end{split}
\]
where $\d E_S^{(2,0)}(v)$, $\d E_S^{(0,2)}(w)$ and $\d E_B^{(0,2)}(w)$ are given by \eqref{eq:dE_buck} and
\[
\begin{aligned}
& \d E_S^{(1,2)}(v,w) = \half\int_\calS\calA^{\a\b\g\d} g_{\b\g} (\nabla_\a v^\g) (\nabla_\g w)(\nabla_\d w) \,\SurfaceElement \\
& \d E_S^{(0,4)}(w) =  \frac{1}{8}\int_\calS \calA^{\a\b\g\d}
(\nabla_\a w)(\nabla_\b w)(\nabla_\g w)(\nabla_\d w)\,\SurfaceElement  \\
& \d E_B^{(1,2)}(v,w) =  - \frac{1}{12} \int_\calS \calA^{\a\b\g\d}  (\nabla_\b\nabla_\a w) (\nabla_\d\nabla_\g v^\eta)(\nabla_\eta w) \,\SurfaceElement \\
& \d E_B^{(0,4)}(w) =  - \frac{1}{24} \int_\calS \calA^{\a\b\g\d} (g^{-1})^{\eta\e} (\nabla_\eta w) (\nabla_\e w)
(\nabla_\b\nabla_\a w)  (\nabla_\d\nabla_\g w)\,\SurfaceElement.
\end{aligned}
\]
We are seeking the perturbation that minimizes the energy variation for small $\e>0$. Since we expect, to leading order, the minimizer to be proportional to $\tw$ (the least stable out-of-plane mode at $\tc$) with a pre-factor that vanishes as $\e\to0$,  we expand the minimizer in a power series in $\e$, whose first terms are
\begin{equation}\label{eq:tildes}
\begin{aligned}
v^\g &=c^2 \tv^\g \,\e^p  + \dots \\
w &=  c \tw \,\e^q + \tilde{\tilde{w}} \,\e^r  + \dots,
\end{aligned}
\end{equation}
where the exponents $p,q,r$ and the constant $c$ are yet to be determined. Substituting this expansion into the energy variation we get
\[
\begin{split}
\d E &=  - c^2   \Brk{\d E_B^{(0,2)}(\tw)} \,\e^{2q+1}  \\
&+ c^4  \Brk{\d E_S^{(2,0)}(\tv)}\,\e^{2p} \\
&+  c^4 \Brk{\d E_S^{(1,2)}(\tv,\tw) + \tc^2\, \d E_B^{(1,2)}(\tv,\tw)} \,\e^{p+2q}  \\
&+ c^4 \Brk{\d E_S^{(0,4)}(\tw) + \tc^2\, \d E_B^{(0,4)}(\tw)}\,\e^{4q}  \\
&+ \O(\e^{2r},\e^{3p},\e^{5q},\e^{2p+q},\e^{p+3q},\e^{2p+1},\e^{4q+1},\e^{p+2q+1})\\
\end{split}
\]
The first term on the right hand side is the quadratic out-of-plane term, which is, as expected, negative for $\e>0$.
For a super-critical bifurcation,
It is balanced by the quartic term, from which we infer that $2q+1=4q$, i.e., $q=1/2$. Since the term that is quadratic in $v$ is positive, it follows that $2p\ge 2q+1=2$. We may then set $p=1$, with the possible outcome that we obtain $\tv=0$ (i.e., that the $v$ terms are sub-dominant).

We proceed to minimize this expression (with all four terms of order $\e^2$) with respect to the in-plane perturbation $\tv$ and the constant $c$. Note that under the normalization choice in \eqref{eq:tildes} the minimizing $\tv$ does not depend on $c$, since the two $\tv$-dependent terms are proportional to $c^4$. The $\tv$-dependent terms consist of a positive-definite quadratic term and a linear term, which guarantees the existence of a non-trivial minimizer (and in particular confirms that $p=1$). Once $\tv$ has been determined, a minimizing $c$ exists if and only if the sum of the terms proportional to $c^4$ are positive. Then,
\begin{widetext}
\[
c^2 = \frac{\half\,\d E_B^{(0,2)}(\tw)}{\d E_S^{(2,0)}(\tv) +
\Brk{\d E_S^{(1,2)}(\tv,\tw) + \tc^2\, \d E_B^{(1,2)}(\tv,\tw)} +
\Brk{\d E_S^{(0,4)}(\tw) + \tc^2\, \d E_B^{(0,4)}(\tw)}}.
\]
\end{widetext}
Recall that $\e c$ is, to leading order in $\e$,  the $L^2$-norm of the out-of-plane deflection $w$.
If the denominator is negative, then the bifurcation is sub-critical and the branch of stable buckled solutions cannot be found by a local analysis about the plane-stress solution.

We calculated $c$ for both the elliptic and flat geometries; recall that in both cases the critical mode is axisymmetric, $n=0$. For the elliptic geometry the bifurcation was found to be super-critical for the whole possible range of curvatures $K$ (for large enough $K$ the surface is no longer an embedding, as the sphere closes upon itself). For the flat geometry a transition from super-critical to sub-critical bifurcations was found: the bifurcation is super-critical for $\alpha$ in the range  $0.58< \alpha < 1$, and sub-critical for $\alpha<0.58$.

\begin{figure}
\begin{center}
\includegraphics[height = 1.25 in]{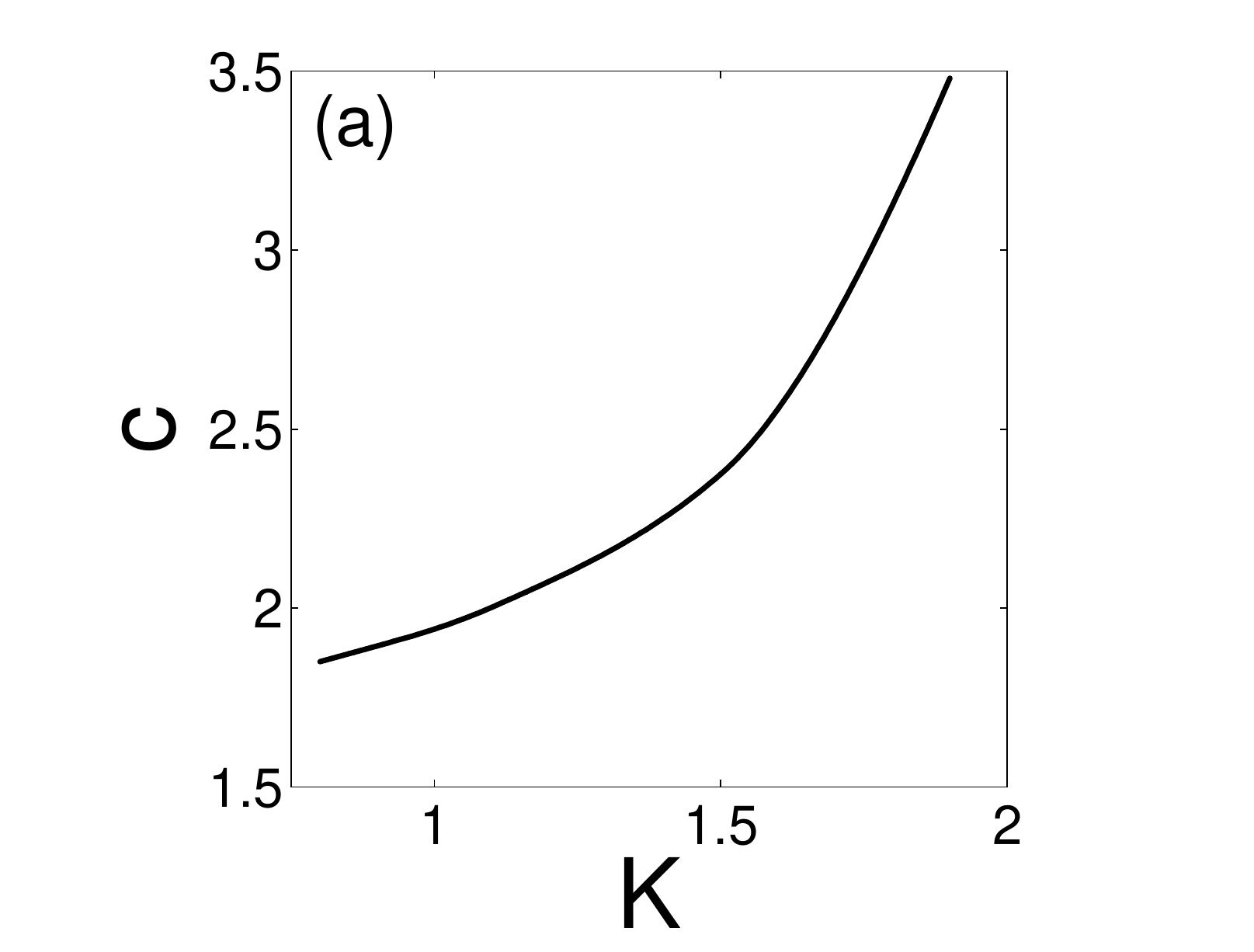}
\includegraphics[height = 1.25 in]{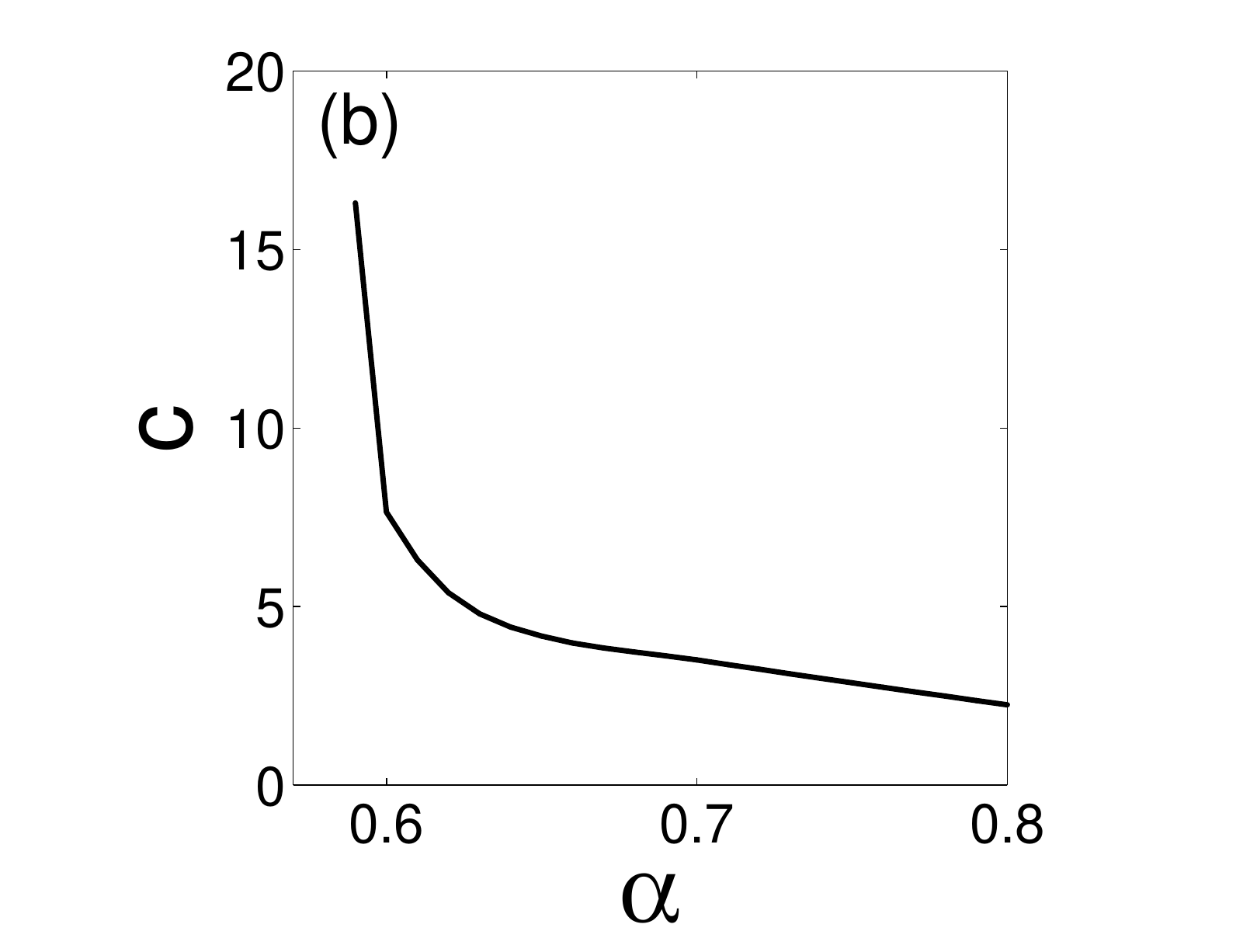}
\end{center}
\caption{The coefficient $c$ versus the Gaussian curvature $K$ of the elliptic geometry (a) and the parameter $\alpha$ of the flat geometry (b).}
\label{fig:bif}
\end{figure}

In \figref{fig:bif} we show the value of $c$ versus the Gaussian curvature $K$ of the elliptic geometry (left) and the parameter $\alpha$ of the flat geometry (right). Note that at the transition point from super-critical to sub-critical bifurcation, $\alpha\approx0.585$, the coefficient $c$ diverges.

\section{Boundary layers in very thin plates}

It was shown in Section~\ref{sec:limit} that provided that a limit equilibrium configuration as $\thickness\to0$ exists, it is given by the isometric immersion that minimizes the Willmore functional.
How a sequence of equilibrium configurations approaches the Willmore isometry is non-trivial.
The convergence is in the Sobolev space $W^{2,2}$---the space of surfaces with square integrable second (weak) derivatives \cite{Ada75}. This guarantees (by the Sobolev embedding theorem) uniform convergence in the space of once-differentiable embeddings, but not in the space of twice-differentiable embeddings. In other words, second derivatives may not converge uniformly.


Almost a hundred years ago (see \cite{FW55} and references therein), it was observed that thin elastic bodies may exhibit boundary layers, which interpolate between a state of minimum stretching content in the bulk and the zero normal traction and zero bending moment conditions at the boundary.  Such boundary layers also occur in non-Euclidean plates, and turn out to dominate the deviation from an isometry as $\thickness\to0$.

Generally speaking, a large thickness implies a bending energy-dominated configuration (i.e., close to flat), whereas a small thickness implies a stretching energy-dominated configuration (i.e., close to an isometry). Whether a thickness $\thickness$ is to be considered as  ``large" or ``small" is determined by comparison with the shortest lengthscale of the problem, which may vary with position. 
For every finite $\thickness$ there exists a distance from the boundary, $\ell$, with respect to which  $\thickness$ cannot be considered small. As a result, we expect bending energy-dominated behavior in a strip of thickness $\ell$ near the boundary. 

We start with a scaling argument.
Let  $h_{\a\b}$ be the second fundamental form of an isometric immersion $\f(x^1,x^2)$ that minimizes the Willmore functional (the metric is of course equal to the reference metric $g_{\a\b}=\go_{\a\b}$). From the point of view of the bending energy  it would be favorable to have a flat surface, $h_{\a\b}=0$, however the second fundamental form cannot be modified without a modification of the metric, as the two must satisfy the Gauss-Mainardi-Codazzi equations. In particular, the Gaussian curvature is an isometric invariant. From the analysis in Appendix~\ref{sec:app},
\emph{assuming that the perturbations do not involve small-scale features}, 
we see that the variation in stretching content is quadratic in the perturbation fields $v,w$, whereas the variation in bending content is linear in $v,w$, i.e.,
\[
\delta E \sim \O(v^2,w^2) + \thickness^2\, \O(v,w).
\] 
Since equilibrium is obtained by a balance of the negative bending contribution and the positive stretching contribution, then $v,w\sim \O(\thickness^2)$, and $\delta E \sim \O(\thickness^4)$. 

These are however bulk considerations, where energy balance is considered uniformly over the surface. The question is whether the total elastic energy can be reduced by a larger, yet \emph{local} change in the bending content density. This is not possible inside the domain, because even a local change in the second fundamental form involves a non-local change in the metric, hence the gain in stretching energy exceeds the loss in bending energy. The situation may however be different at the boundary.

In a bending-dominated region we expect the curvatures to deviate from the curvatures associated with the Willmore isometry by $\O(1)$. As curvatures relate to the metric through two differentiations, such a deviation over a strip of width $\thickness^p$, 
induces a metric deviation of $\O(t^{2p})$. Thus, the variation in total energy is of order
\[
\delta E \sim \O(\thickness^{4p}) + \thickness^2 \, \O(1),
\]
which yields, $p=1/2$. 

To make this into a rigorous argument, 
suppose that $v=\O(\thickness^q)$ and  $w=\O(\thickness^r)$, where $q,r>0$, both varying over a boundary layer whose width scales like $\thickness^p$, where $p>0$. Inside the boundary layer, the variation in stretching energy density \eqref{eq:dwS_full} is dominated by terms of order
\[
\d w_S = \O(\thickness^{2q-2p},\thickness^{2r},\thickness^{q+r-p}
,\thickness^{q+2r-3p},\thickness^{3r-2p},\thickness^{4r-4p}).
\]
Note that this contribution is always positive. On the other hand, the variation in bending energy density, \eqref{eq:dwB_full}, which can become negative, is dominated for small $t$ by terms of order
\[
\thickness^2\,\d w_B = \O(\thickness^{q-p+2},\thickness^{r-2p+2}).
\]
The exponents $q,r,p$ are determined such to maximize the change in (negative) bending energy, without the (positive) stretching energy becoming dominant. That is, if we define
\[
\begin{aligned}
e_S &= \min(2q-2p, 2r, q+r-p, q+2r-3p, 3r-2p, 4r-4p) \\
e_B &= \min(q-p+2,r-2p+2),
\end{aligned}
\]
then we need to choose $q,r,p$ such to minimize $e_B$ subject to the constraint that $e_S\ge e_B$. It can be shown that the optimal choice satisfies  $r=1$, $p=1/2$, and $q\ge 3/2$. That is, the width of the boundary layer is expected to scale like the square root of the plate thickness. To minimize the gain in stretching content, we expect a balance between the $\nabla_\a v^\b$ and $w$ terms, which yields $q=3/2$. 

\begin{figure}
\putfig{2.5}{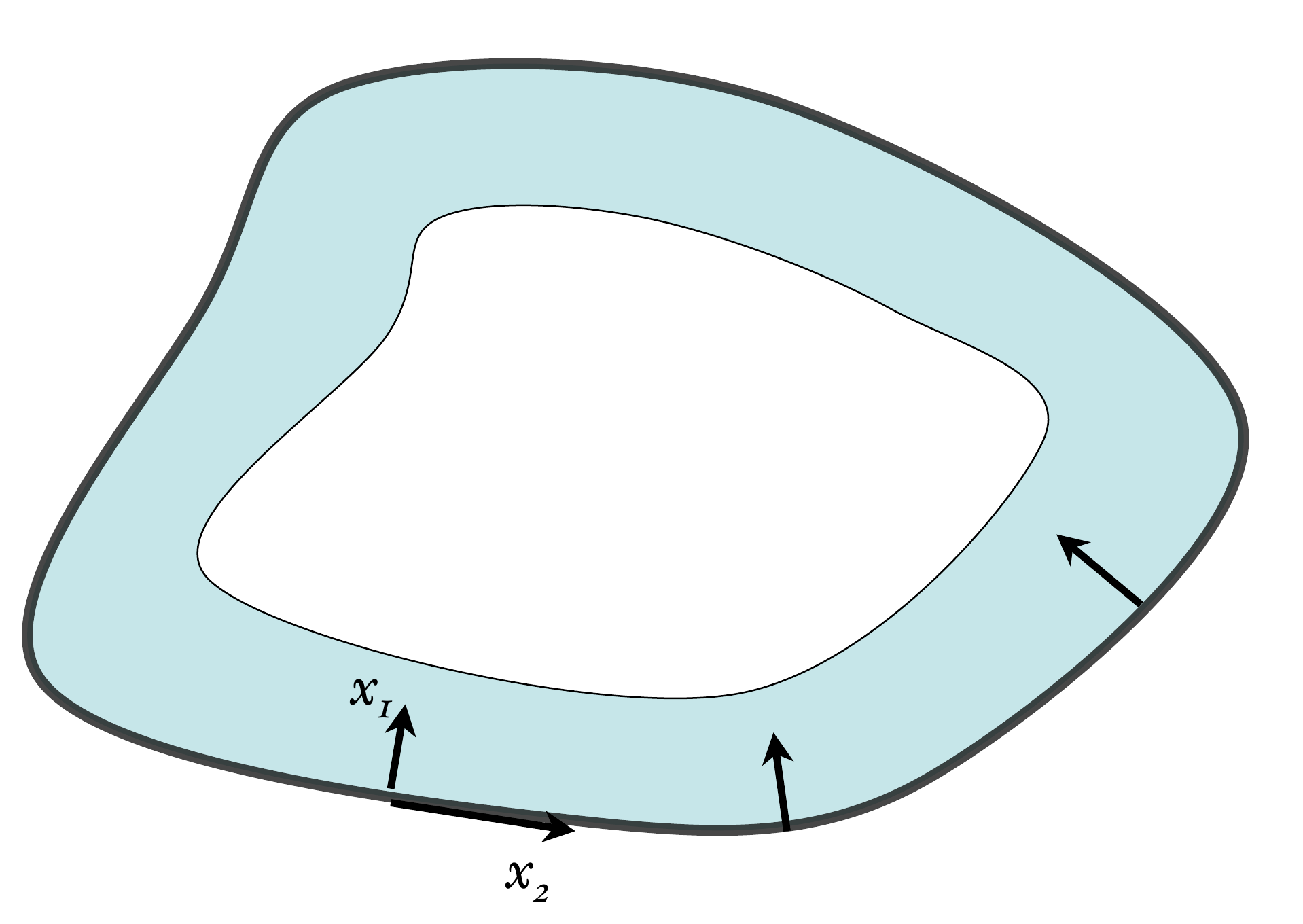}
\caption{(Color online) Local parametrization of an annulus bounded by a boundary of the domain.}
\label{fig:AnnularStrip}
\end{figure}

Assuming these scaling exponents, we study the structure of the boundary layer. 
We consider, as before, a perturbation, which we decompose as
\[
\d\f = v^\g \,\df\g + w\N.
\]
Consider now a local parametrization $\f:[0,\ell^1)\times[0,\ell^2)\to\R^3$ of the surface, such that the parametric line $x^1=0$ coincides with a boundary of the surface, with the positive $x^1$ axis inside the sample. Moreover, the parametrization  of the unperturbed surface
is  semi-geodesic, i.e., $g_{11}=1$, $g_{12}=g_{21}=0$ and  $g_{22}=\phi^2$; such a parametrization is always possible. One may also set
$g_{22}=1$  along the boundary (see \figref{fig:AnnularStrip}).

Since we expect a boundary layer of size $\sqrt{\thickness}$,
we stretch the positive $x^1$ axis accordingly by introducing a rescaled coordinate,
\[
X^1 = \frac{x^1}{\sqrt{\thickness}},
\]
and rescale the perturbations $v^\a,w$, such that the new variables \emph{and their derivatives} are all of order one,
\beq
\begin{aligned}
V^\g(X^1,x^2) &= \frac{1}{\thickness^{3/2}} v^\g\brk{\sqrt{\thickness} X^1,x^2} \\
W(X^1,x^2) &= \frac{1}{\thickness} w\brk{\sqrt{\thickness} X^1,x^2}.
\end{aligned}
\label{eq:VW}
\eeq
By setting, say, $\ell^1\sim \O(\thickness^{1/4})$ we have a situation where, as $\thickness\to0$, the local coordinates $(x^1,x^2)$ parametrize a shrinking annulus which converges to the boundary, whereas in the rescaled coordinates, $(X^1,x^2)$, the range of $X^1$ in the positive direction tends to infinity. We are going to show the existence of a perturbation of such structure that reduces the total elastic energy.

We then evaluate the variation in energy content densities inside the boundary layer, i.e., at points $x^1 = \sqrt{\thickness}\,X^1$, with $X^1\sim \O(1)$. To leading order, the unperturbed metric and the Christoffel symbols are given by their values at the boundary, and covariant derivatives coincide with partial derivatives. Since $\go_{\a\b}=g_{\a\b}$ we also have
\beq
\calA^{\a\b\g\d} = \frac{\nu}{1-\nu} \d_{\a\b} \d_{\g\d} + \half\brk{\d_{\a\g} \d_{\b\d} +  \d_{\a\d} \d_{\b\g}} + \O(\thickness^{1/2}).
\label{eq:calA_approx}
\eeq

Substituting the rescaled variables \eqref{eq:VW} into the variations \eqref{eq:dwS_full} and \eqref{eq:dwB_full} in stretching and bending content densities, we get
\[
\begin{aligned}
\d w_S/\thickness^2 &=
\half \calA^{1\b1\d}   (\partial_1 V^\b) (\partial_1 V^\d) +
\half \calA^{\a\b\g\d} h_{\a\b} h_{\g\d} W^2\\
&- \calA^{1\b\g\d} h_{\g\d}  (\partial_1 V^\b) W
+ \half\calA^{1111}  (\partial_1 V^1)  (\partial_1 W)^2 \\
&- \half \calA^{\a\b11} h_{\a\b}  (\partial_1 W)^2 W
+ \frac{1}{8}\calA^{1111} (\partial_1 W)^4 + \O(\thickness^{1/2}),
\end{aligned}
\]
and
\[
\begin{aligned}
\d w_B &=  \frac{1}{12} \calA^{\a\b11}  h_{\a\b} (\partial_1\partial_1 W) +
\frac{1}{24} \calA^{1111}  (\partial_1\partial_1 W)^2  + \O(\thickness^{1/2}).
\end{aligned}
\]
Substituting also expression \eqref{eq:calA_approx} for the elastic tensor, we end up with
\[
\begin{aligned}
\d w_S/\thickness^2  &=
\frac{1}{4}\Brk{(\partial_1 V^2) - 2 h_{12} W}^2 + \frac{1}{2} h_{22}^2 W^2  \\
&+ \frac{\nu}{8(1-\nu)} \Brk{2 (\partial_1 V^1) - 2 (h_{11}+h_{22}) W + (\partial_1 W)^2}^2  \\
&+ \frac{1}{8} \Brk{2 (\partial_1 V^1)  - 2 h_{11} W   + (\partial_1 W)^2}^2 + \O(\thickness^{1/2}) \\
\d w_B  &=
\frac{1}{24(1-\nu)}\Brk{2 \brk{h_{11}+\nu h_{22}} (\partial_1\partial_1 W)
+  (\partial_1\partial_1 W)^2} \\
&+ \O(\thickness^{1/2}).
\end{aligned}
\]

The variation in stretching content density is a sum of squares. The optimal choice of $V^2$ is the one that makes the first term vanish, namely,
\[
(\partial_1 V^2) = 2W h_{12}.
\]
Similarly, the optimal choice for $V^1$ satisfies
\[
2 (\partial_1 V^1) + (\partial_1 W)^2 = 2\brk{h_{11} + \nu h_{22}} W,
\]
so that the variational problem reduces into one for $W$ only,
\[
\left.\frac{\d w_S}{\thickness^2}\right|_{V^1,V^2\text{ optimal}}  =  \half(1+\nu) h_{22}^2 W^2 + \O(\thickness^{1/2}).
\]
Omitting the $\O(\thickness^{1/2})$ terms, 
the resulting Euler-Lagrange equations are
\[
\frac{\partial^4 W}{\partial (X^1)^4} + 12(1-\nu^2)\, h_{22}^2 W  = 0,
\]
with boundary conditions
\[
\frac{\partial^2 W}{\partial (X^1)^2} = -\brk{h_{11} + \nu h_{22}},
\Textand
\frac{\partial^3 W}{\partial (X^1)^3} = 0,
\]
at $X^1=0$, and  $W(\infty,x^2)=0$.
The solution is
\beq
W(X^1,2) = -\frac{h_{11} + \nu h_{22}}{2\lambda^2}  e^{-\lambda X^1} \brk{\cos \lambda X^1-\sin \lambda X^1},
\label{eq:BLsol}
\eeq
where
\[
\lambda = \Brk{3 (1-\nu^2) h_{22}^2}^{1/4}.
\]
Reverting to the original unscaled units, $(x^1,x^2$), the out-of-plane perturbation $w(x^1,x^2)$ exhibits a boundary layer whose width $\ellBL$ is
\[
\ellBL =  \Brk{3 (1-\nu^2)}^{-1/4} \sqrt{\thickness/|h_{22}|}.
\]

Substituting the asymptotic boundary layer profile \eqref{eq:BLsol} in the leading order expressions for $\d w_S/\thickness^2$ and $\d w_B$, we get
\beq
\begin{aligned}
\d w_S/\thickness^2 &=
\frac{(h_{11} + \nu h_{22})^2}{24(1-\nu)}
e^{-2\lambda X^1} \brk{\cos \lambda X^1-\sin \lambda X^1}^2 \\
\d w_B &=
\frac{(h_{11} + \nu h_{22})^2}{24(1-\nu)}
e^{-2\lambda X^1} \brk{\cos \lambda X^1+\sin \lambda X^1}\\
&\times
\brk{\cos \lambda X^1+\sin \lambda X^1 -2\, e^{\lambda X^1}}.
\end{aligned}
\label{eq:BL_wSwB}
\eeq

At the boundary, to leading order, curvature in the direction normal to the boundary is given by 
\[
h_{11}+ ֿ\partial_1 \partial_1 w = -\nu h_{22}.
\]
This leads to the vanishing of the bending moment at the boundary, i.e., $m^{11}=0$, which is one of the boundary conditions associated with the energy minimization variational problem \cite{ESK08}. 
Note that the normal bending moment along the boundary is proportional to $h_{11}+\nu h_{22}$. 
If $\nu=0$ and the bending minimizing isometry satisfies $h_{11}=0$, then no correction is needed in order to satisfy the boundary conditions, and we expect no boundary layer to develop. This fact is manifested by the vanishing of $W$ in \eqref{eq:BLsol}. 
For $\nu\ne0$, however, the curvatures normal and tangent to the boundary have opposite  signs, which means that the surface is hyperbolic in the vicinity of the boundary. Finally, our analysis breaks down in the event that $h_{22}=0$, in which case a boundary layer of different nature may emerge.

\begin{figure}
\begin{center}
\includegraphics[height=2.75in]{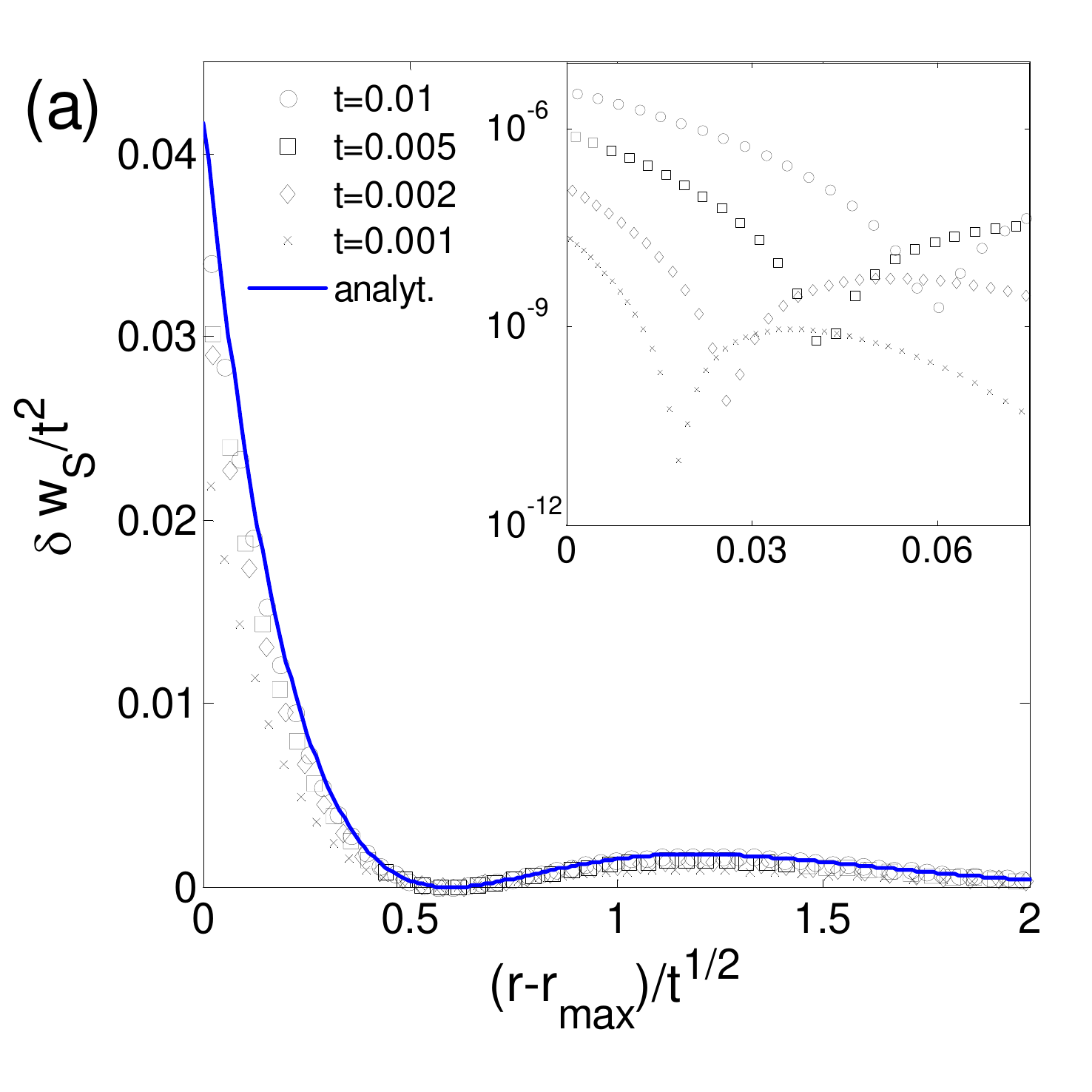}
\includegraphics[height=2.75in]{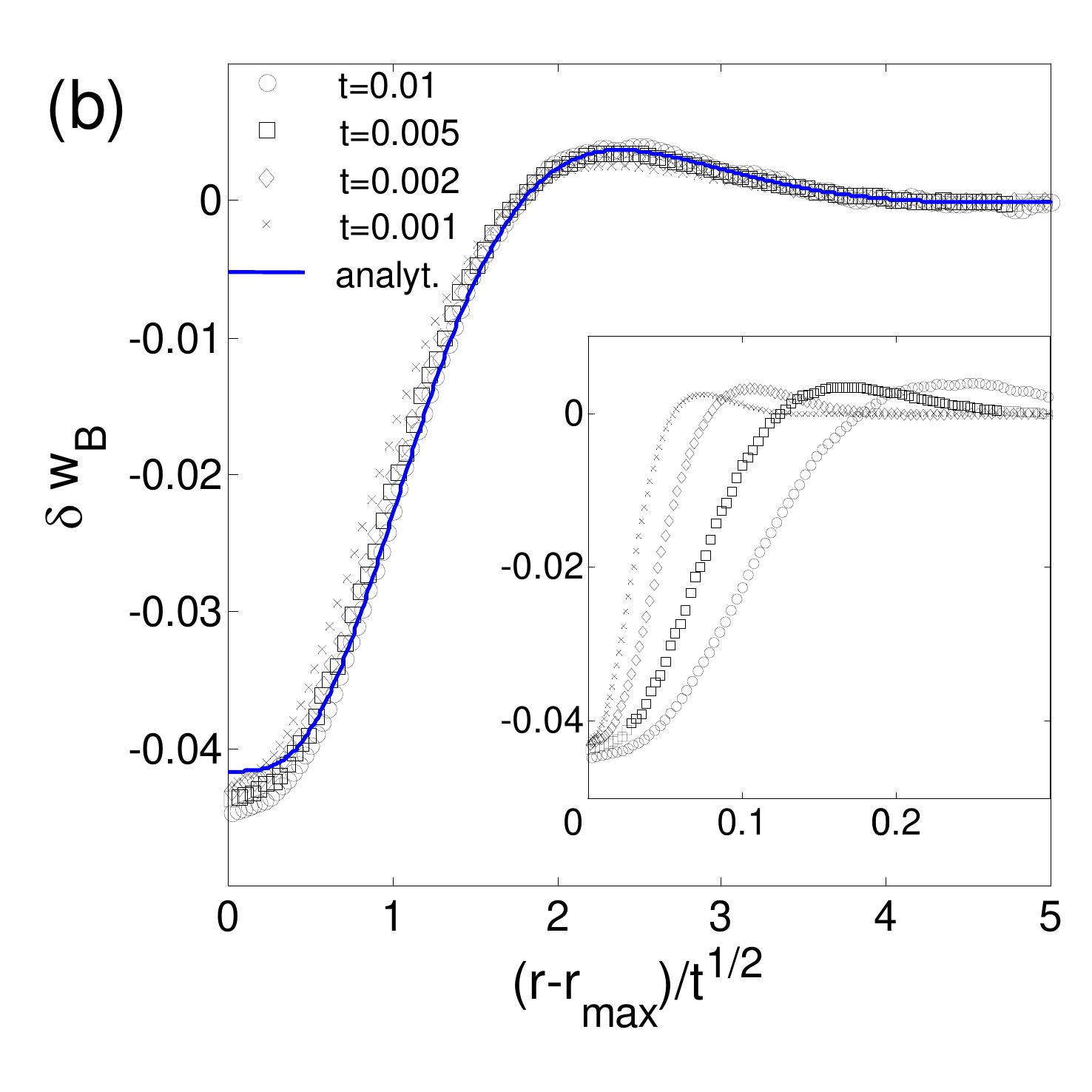}
\end{center}
\caption{(Color online) Structure of the boundary layer near the outer boundary, $r=1.1$, for the elliptic geometry \eqref{eq:elliptic} and the same parameters as in previous sections. The  top figure (a) shows the rescaled deviation in stretching content density, $\d w_S/\thickness^2$ versus the rescaled coordinate $(\rmax-r)/\sqrt{\thickness}$, calculated for four different thicknesses, $h=0.01, 0.005, 0.002, 0.001$ (symbols). The solid (blue) line is the asymptotic result  \eqref{eq:BL_wSwB}. The inset shows the unscaled stretching content density deviations versus the unscaled coordinate $\rmax-r$. Note the logarithmic $y$-scale. The bottom figure (b) shows the bending content density, $\d w_B$, versus the rescaled coordinate for the same values of the thickness. The inset shows the unscaled results. }
\label{fig:BL}
\end{figure}

To validate our results, we plot in \figref{fig:BL} the rescaled deviations in stretching content density, $\d w_S/\thickness^2$, and the deviations in bending content density, $\d w_B$, versus the rescaled coordinate $(\rmax-r)/\sqrt{\thickness}$, for the elliptic geometry \eqref{eq:elliptic}, with the same parameters as in previous sections, namely, $\rmin=0.1$, $\rmax=1.1$, $K=1$, and $\nu=0$. The results were obtained by minimizing the full energy functional. We display results for $\thickness=0.01$, $0.005$, $0.002$, and $0.001$. As expected, the four rescaled curves approximately coincide. These
numerically computed curves are compared with our asymptotic expressions \eqref{eq:BL_wSwB}. 

\section{Discussion}

A theory of non-Euclidean plates, applicable to thin elastic sheets that do not have a stress-free rest configuration, has recently been proposed in \cite{ESK08}. This paper provides a first mathematical analysis of this model. Two different limits of such plates are analyzed: (i) the buckling transition, and (ii) the occurrence of boundary layers in the limit where the plate thickness tends to zero, and the configuration is expected to converge to the isometric immersion that minimizes the Willmore functional.

We proved a general result, whereby any non-Euclidean plate that does not have a flat stress-free configuration (i.e., whose reference metric has non-zero Gaussian curvature), buckles if the plate is sufficiently thin. The transition from flat to buckled equilibria may be either continuous, or discontinuous, depending on the particular system. Instances of both types have been observed.

We showed that in the thin plate regime, the dominant deviation from the Willmore isometry is governed by a bending-dominated boundary layer, whose structure was calculated using a boundary layer analysis.
In particular, the width of this boundary layer is determined by both the plate thickness and the  tangential curvature at the boundary of the Willmore isometry---it scales like the square root of their product. It would be of interest to observe the occurrence of such boundary layers in experiments, e.g, in the thermo-responsive gels studied in \cite{KES07}, in order to further validate the model as describing thin elastic sheets  with no stress-free configuration.

The reported results are of high relevance to shape formation in growing biological tissues. Spontaneous buckling and wrinkling of non-Euclidean sheets was suggested as a mechanism for shaping leaves that are free of external confinement \cite{NCCC03,SMS04,SRS07,DB08}. These studies were either qualitative, or assumed the limit of an infinitely-thin sheet. As such, their results are relevant  only to selected cases. Recent studies, suggest and demonstrate that the mechanical stress field can lead to differentiation of cells \cite{Far03,HHJKUBCSBMCT08,LCCBLMJ08}, and might act as a regulator of tissue growth \cite{Shr05}. These studies emphasize the need to know the mechanical state of a tissue of finite thickness, which  undergoes differential growth. Our calculations of buckling threshold, as well as the pre- and post-buckling stress distribution within plates, can be integrated into the biological picture as inputs that can affect its future  evolution. In particular, the existence of a boundary layer and its scaling predict thickness-dependent, localized effects near the  boundaries of unconstrained growing tissues.


\begin{acknowledgments}
This work was supported by the United States–Israel Binational Foundation (grant no. 2004037)
and the MechPlant project of European Union’s New and Emerging Science and Technology program.
We thank Yossi Shamai for many useful discussions. RK is grateful to the Department of Chemical  and Biomolecular Engineering at Rice University for its hospitality.
\end{acknowledgments}

\appendix

\section{The $\thickness\to0$ limit}
\label{app:gamma}

The $\thickness\to0$ limit can be approached in two way. The first possibility is to depart from the three-dimensional model, i.e., the energy functional \eqref{eq:3Denergy}, and analyze the limit of the energy minimizers (or approximate energy minimizers) as $\thickness\to0$. To this end, one would hope to be able to use the analytical techniques based on $\Gamma$-convergence developed in \cite{FJM02a,FJM02b,FJM06}. There is however an obstacle that prevents the direct application of the abovementioned techniques to the present context: we do not have a reference configuration with respect to which deviations can be measured. Indeed, the analysis in \cite{FJM02b} relies heavily on a so-called rigidity property that estimates the distance of the displacement from a rigid rotation in terms of an integral over local distances from rotations.

The second alternative is to depart from the two-dimensional model, i.e., the energy functional \eqref{eq:Efunc}.  For reasons to be clarified below, we work with an energy functional rescaled with the thickness square,
\beq
\Ft= \frac{E}{\thickness^2} = \frac{1}{\thickness^2} E_S + E_B,
\label{eq:Ft}
\eeq
where the notation $\Ft$ makes the $\thickness$-dependence explicit.
Clearly, for every fixed $\thickness$ the functionals $E$ and $\Ft$ have the same minimizers.
We view $\Ft$ as a one-parameter family of functionals, defined on the Sobolev space $W^{2,2}(\calS;\R^3)$, of surfaces whose second (weak) derivatives are in $L^2(\calS)$. Since we view every two configurations that differ by a rigid motion as identical, the space of immersions is in fact the quotient space $W^{2,2}(\calS;\R^3)$ modulo rigid motions. We denote by $\Ft[\f]$, $E_S[\f]$, and $E_B[\f]$ the functionals $\Ft$, $E_S$, and $E_B$ evaluated at a configuration $\f=\f(x^1,x^2)$. We will also denote by
\[
e_\thickness = \inf\BRK{\Ft[\f]:\,\, \f\in W^{2,2}(\calS;\R^3)}
\]
the $\thickness$-dependent greatest lower bounds on the energy.

The two-dimensional elastic problem formulated in the Section~\ref{sec:model} assumes the existence of a family of minimizers, $\ft^*$, such that $\Ft[\ft^*] = e_\thickness$. Even if such minimizers do not exist, we can always construct a family of \emph{approximate minimizers}, $\ft^*$, satisfying,
\[
\lim_{\thickness\to0} \brk{\Ft[\ft^*] - e_\thickness} = 0.
\]

Suppose now that the two-dimensional reference metric $\go_{\a\b}$ assumes an isometric immersion $\f= \hat{\f}$ with finite bending content.  Then for every $\thickness$,
\[
e_\thickness \le \Ft[\hat{\f}] = E_B[\hat{\f}],
\]
i.e., the greatest lower bounds on the energy are uniformly bounded. In particular, it follows that
\[
\lim_{\thickness\to0} E_S[\ft^*] = 0,
\]
which means that the metrics $g(\ft^*)$ associated with the family of approximate minimizers converge, in least-square norm, to the  reference metric $\go$.

The mean-square convergence of the family of metrics, as $\thickness\to0$, does not guarantee that the family of (possibly approximately) minimizing configurations has a limit  (modulo rigid motions).  If, however, such a limit does exist, then we show that this limit is an isometric immersion that minimizes the bending content, i.e., the Willmore functional. Specifically,

\begin{quote}
\emph{Let $\go$ be a reference metric that assumes a $W^{2,2}$ isometric immersion. Let $\ft^*$ be a family of approximate minimizers of the functionals $\Ft$. If the family $\ft^*$ (weakly) converges in $W^{2,2}(\calS;\R^3)$, as $\thickness\to0$, to a limit $\f^*$, then $\f^*$ is a configuration that minimizes the Willmore functional over all isometric immersions of the reference metric $\go$.}
\end{quote}

To prove this theorem we construct a ``limit functional",
\[
F_0[\f] = \begin{cases}
E_B[\f] & g(\f) = \go \\ \infty & \text{otherwise},
\end{cases}
\]
and show that the functionals $\Ft$ $\Gamma$-converge to $F_0$, as $\thickness\to0$, with respect to the weak $W^{2,2}$ topology. It then follows that every converging sequence of approximate minimizers of $\Ft$ converges to a minimizer of $F_0$ \cite{Dal93}.

To show that $\Ft$ $\Gamma$-converges to $F_0$ we need to show that:
\begin{enumerate}
\item Lower-semicontinuity: for every sequence $\ft$ that converges to a configuration $\f$ (in the weak $W^{2,2}$ topology),
\[
\liminf_{\thickness\to0} \Ft[\ft] \ge F_0[\f].
\]
\item Recovery sequence: for every $\f\in W^{2,2}$ there exists a sequence $\ft$ that weakly converges to $\f$ for which
\[
\liminf_{\thickness\to0} \Ft[\ft] = F_0[\f].
\]
\end{enumerate}

To prove the lower-semicontinuity property we note that the weak $W^{2,2}$ convergence of $\ft$ to $\f$ implies the weak convergence of the corresponding metrics, $g(\ft)\to g(\f)$, in the weak $W^{1,2}$ topology, which by the Sobolev embedding theorem \cite{Ada75} implies the convergence of the metrics in the strong $C^0$ topology, i.e., uniform convergence. It follows at once that  the corresponding family of second fundamental forms weakly converges in $L^2$ to the second fundamental form of $\f$, $h(\ft)\to h(\f)$. Since the bending content is equivalent to an $L^2$ norm of the second fundamental form, it follows at once that
\[
\liminf_{\thickness\to0} E_B[\ft] \ge E_B[\f].
\]
Now either $g(\f) = \go$, in which case
\[
\lim_{\thickness\to0} \Ft[\ft] \ge \liminf_{\thickness\to0} E_B[\ft] \ge E_B[\f] = F_0[\f],
\]
or $g(\f) \ne \go$, in which case
\[
\liminf_{\thickness\to0} \Ft[\ft] = \infty = F_0[\f].
\]

To prove the existence of a recovery sequence we take, given $\f$, the constant sequence, $\ft=\f$.
If $g(\f) = \go$ then
\[
\lim_{\thickness\to0} \Ft[\ft] = E_B[\f] = F_0[\f].
\]
If, however, $g(\f) \ne \go$ then
\[
\lim_{\thickness\to0} \Ft[\ft]  = \infty = F_0[\f].
\]

\parag{Comments}
\begin{enumerate}
\item The assumption that the reference metric can be embedded isometrically with finite bending is by no means trivial. The Nash-Kuiper embedding theorem only guarantees the existence of a $C^1$ embedding. Embeddings of class $W^{2,2}$ have been shown to exist under additional assumptions (see, e.g., \cite{HH06}), however there is no general existence proof for arbitrary metrics.

\item We use the weak $W^{2,2}$ topology because we aim to eventually prove that  \emph{every} family of approximate minimizers has a converging subsequence (implying that the limit is a minimizer of the Willmore functional). Such a compactness result cannot possible hold in the strong $W^{2,2}$ topology.

\item A side-result of the above theorem is that the Willmore functional has a  minimizer (although not necessarily unique), even if the functionals $\Ft$ do not have minimizers.
\end{enumerate}

\section{Perturbation analysis}
\label{sec:app}

Consider a sufficiently regular surface $\f(x^1,x^2)$. Any small enough perturbation  can be decomposed into a sum of in-plane and out-of-plane displacements,
\beq
\delta\f = v^\g \,\df\g + w\,\N,
\label{eq:df}
\eeq
where $\N$ is the unit vector normal to the surface. Given such a perturbation we are going to calculate the variation in the elastic energy.

In order to retain the tensorial nature of the problem (coordinate invariance), we need to only utilize covariant differentiation. To do so we need to specify a metric with respect to which the Christoffel symbols are defined. It turns out that choosing the (natural) induced metric on the surface, yields the most compact form for the variation in energy.

We recall the definitions of the covariant derivatives of a scalar field $W$, a contravariant vector $V^\g$, a covariant vectors $V_\g$, and a mixed tensor $T^\b_\g$,
\beq
\begin{aligned}
\nabla_\a W &= \partial_\a W \\
\nabla_\a V^\g &= \partial_\a V^\g + \Chr\g\a\b V^\b \\
\nabla_\a V_\g &= \partial_\a V_\g - \Chr\b\a\g V_\b \\
\nabla_\a T^\b_\g &= \partial_\a T^\b_\g + \Chr\b\a\d T^\d_\g - \Chr\d\a\g T^\b_\d.
\end{aligned}
\label{eq:cov_der}
\eeq
Note that $\nabla_\a$ and $\nabla_\b$ commute only when operating on scalars. As operators on higher rank tensors their commutator is nonzero, and relates to the Gaussian curvature of the surface.

To calculate the variation in energy we need to calculate the variation in the two  fundamental forms.
For this we use the  Gauss-Weingarten equations,
\[
\partial_\a \partial_\b \f =\Gamma^\g_{\a\b}\partial_\g \f+\ h_{\a\b}\N
\Textand
\partial_\a\N= -T^\b_\a \,\partial_\b\f,
\]
where $T^\b_\a = (g^{-1})^{\b\g} h_{\g\a}$. (Note that by our definitions of index raising, $T^\b_\a = h^\b_\a$ only if $g_{\a\b} = \go_{\a\b}$.)
It follows that for a vector in $\R^3$ in the form
\[
\bs{v} = a^\a\,\df\a + b\N,
\]
its derivative is given by
\beq
\partial_\b\bs{v} = \brk{\nabla_\b a^\a - b T^\a_\b}\df\a + \brk{\nabla_\b b + a^\a h_{\a\b}}\N.
\label{eq:derivR3}
\eeq

\subsection{Variation in stretching content}

Differentiating \eqref{eq:df} and using \eqref{eq:derivR3} we get
\beq
\partial_\a\delta\f = \Brk{(\nabla_\a v^\g) - w\,T^\g_\a} \partial_\g\f + \Brk{(\nabla_\a w) + v^\g \,h_{\a\g}}\N,
\label{eq:diff_me}
\eeq
from which we derive the variation in the metric,
\beq
\begin{split}
\d g_{\a\b} &=
\partial_\a\f\cdot\partial_\b\d\f+\partial_\a\d\f\cdot \partial_\b\f+\partial_\a\d\f\cdot \partial_\b\d\f\\
&=   g_{\b\g} (\nabla_\a v^\g) + g_{\a\g} (\nabla_\b v^\g) - 2 w h_{\a\b}\\
&\quad+  \Brk{(\nabla_\a w) + v^\g h_{\a\g}} \Brk{(\nabla_\b w) + v^\d h_{\b\d}} \\
&\quad+
\Brk{(\nabla_\a v^\g) - w T^\g_\a} g_{\g\d}
\Brk{(\nabla_\b v^\d) - w T^\d_\b}.
\end{split}
\label{eq:g_pert}
\eeq
Substituting into \eqref{eq:wswb} we obtain the variation in stretching content density,
\beq
\begin{split}
\d w_S &= \half \stress^{\a\b} \d g_{\a\b} + \frac{1}{8}\calA^{\a\b\g\d} \d g_{\a\b} \d g_{\g\d} \\
&= \d w_S^{(1,0)}(v) + \d w_S^{(0,1)}(w) + \d w_S^{(2,0)}(v) + \d w_S^{(0,2)}(w)  \\
&\quad +  \d w_S^{(1,1)}(v,w) + \d w_S^{(1,2)}(v,w) +  \d w_S^{(0,3)}(w) +  \d w_S^{(0,4)}(w) \\
&\quad +  \O(v^3,v^2 w, v w^3, w^5),
\end{split}
\label{eq:dwS_full}
\eeq
where the various $\d w_S^{(i,j)}$ represent terms of different orders in $v$ and $w$,
\[
\begin{aligned}
\d w_S^{(1,0)}(v) &=  \stress^{\a\b} g_{\b\g} (\nabla_\a v^\g) \\
\d w_S^{(0,1)}(w) &=  - \stress^{\a\b} h_{\a\b} w \\
\d w_S^{(2,0)}(v) &=  \half \stress^{\a\b} h_{\a\g} h_{\b\eta} v^\g  v^\eta  +
\half  \stress^{\a\b}  g_{\g\e} (\nabla_\a v^\g)(\nabla_\b v^\e) \\
&\quad + \half \calA^{\a\b\g\d} g_{\b\eta} g_{\d\e}  (\nabla_\a v^\eta) (\nabla_\g v^\e) \\
\d w_S^{(0,2)}(w) &= \half \stress^{\a\b} \Brk{(\nabla_\a w) (\nabla_\b w) +  H_{\a\b} w^2} \\
&\quad+ \half \calA^{\a\b\g\d} h_{\a\b} h_{\g\d} w^2 \\
\d w_S^{(1,1)}(v,w) &= \stress^{\a\b} h_{\b\g}  v^\g  (\nabla_\a w) -
\stress^{\a\b}   h_{\a\eta} (\nabla_\b v^\eta) w \\
&\quad - \calA^{\a\b\g\d} h_{\g\d} g_{\b\eta} (\nabla_\a v^\eta) w   \\
\d w_S^{(1,2)}(v,w) &= \half\calA^{\a\b\g\d} g_{\b\eta} (\nabla_\a v^\eta)
\Brk{(\nabla_\g w)(\nabla_\d w) + H_{\g\d}  w^2} \\
& \quad -\calA^{\a\b\g\d}h_{\a\b}h_{\d\mu}v^\mu w (\nabla_\g w) \\
& \quad +\calA^{\a\b\g\d}h_{\a\b}h_{\d\mu} w^2 (\nabla_\g v^\mu)  \\
\d w_S^{(0,3)}(w) &= -\half \calA^{\a\b\g\d} h_{\a\b}
\Brk{(\nabla_\g w)(\nabla_\d w) + H_{\g\d}  w^2} w  \\
\d w_S^{(0,4)}(w) &= \frac{1}{8}\calA^{\a\b\g\d}
\Brk{(\nabla_\a w)(\nabla_\b w) + H_{\a\b}  w^2}\\
&\qquad\quad\,\, \times
\Brk{(\nabla_\g w)(\nabla_\d w) + H_{\g\d}  w^2},
\end{aligned}
\]
where we have used the symmetry of $s^{\a\b}$ and $\calA^{\a\b\g\d}$,
and  introduced the following new symmetric tensor,
\[
H_{\a\b} = T^\g_\a  h_{\g\b} = (g^{-1})^{\g\d} h_{\d\a}h_{\g\b},
\]
which is known as the third quadratic form.

\parag{Perturbation of a flat surface}
When the unperturbed surface is flat then $h_{\a\b} = H_{\a\b}=0$, which simplifies \eqref{eq:dwS_full} considerably. In particular, all the terms that are odd functions of the out-of-plane perturbation $w$ vanish.

\parag{Perturbation of an isometric immersion}
If the unperturbed surface is an isometric immersion, $g_{\a\b} = \go_{\a\b}$, then $s^{\a\b}=0$, which implies that the lowest-order non-vanishing terms in \eqref{eq:dwS_full} are
\[
\begin{split}
\d w_S &=
\half \calA^{\a\b\g\d} \brk{g_{\b\eta} (\nabla_\a v^\eta) - h_{\a\b} w}
\brk{g_{\g\e} (\nabla_\d v^\e) - h_{\g\d} w} \\
&\quad + \O(v^3,v^2 w, v w^2, w^3).
\end{split}
\]
Note that it is explicitly assumed here that derivatives of the deviations $v,w$ are of the same order of magnitude as the deviation themselves. This assumption breaks down in the presence of  small scale features, such as boundary layers.

\subsection{Variation in bending content}

To calculate the variation in the second quadratic form we start with
\[
h_{\a\b} = \partial_\a\partial_\b \f \cdot\N = -\partial_\b \f \cdot\partial_\a\N.
\]
from which follows that
\beq
\d h_{\a\b} = -\partial_\b \d\f \cdot\partial_\a\N -\partial_\b \f \cdot\partial_\a\d\N
-\partial_\b \d\f \cdot\partial_\a\d\N.
\label{eq:dh1}
\eeq
The first term follows directly from \eqref{eq:diff_me} and the Weingarten equation,
\beq
-\partial_\b \d\f \cdot\partial_\a\N = h_{\a\g} (\nabla_\b v^\g) - w H_{\a\b}.
\label{eq:I}
\eeq

To calculate the second and third terms, we need to express the perturbation $\d\N$ in the unit normal vector. For that, we use the facts that
$\d(\N\cdot\N) = \d(\df\a\cdot\N) = 0$,
hence,
\[
2\N\cdot\d\N + \d\N\cdot\d\N = \partial_\a\d\f\cdot\N + \partial_\a\f\cdot\d\N  + \partial_\a\d\f\cdot\d\N = 0.
\]
Setting
\beq
\d\N = (g^{-1})^{\g\d} a_\d \,\partial_\g\f + b\N,
\label{eq:dN}
\eeq
this yields a closed set of equations for the three coefficients $a_\g,b$,
\beq
\begin{aligned}
a_\a   &= - (1+b)\Brk{(\nabla_\a w)+ v^\g h_{\a\g}} -
a_\b  \Brk{(\nabla_\a v^\b) - w T^\a_\b} \\
b &= -\half (g^{-1})^{\g\b} a_\g a_\b  - \half b^2.
\end{aligned}
\label{eq:ab}
\eeq
Applying \eqref{eq:derivR3} on \eqref{eq:dN} we get
\beq
\begin{split}
\partial_\a\d\N &= \Brk{(g^{-1})^{\g\d} \nabla_\a a_\d - b T^\g_\a} \df\g
+ \Brk{\nabla_\a b + T^\d_\a a_\d}\N,
\end{split}
\label{eq:ddN}
\eeq
where we have used the fact that the covariant derivative of $g_{\a\b}$ and its inverse vanishes. It follows at once that the second term in \eqref{eq:dh1} is given by
\beq
-\partial_\b \f \cdot\partial_\a\d\N = - (\nabla_\a a_\b) + b h_{\a\b}.
\label{eq:II}
\eeq
The third term in \eqref{eq:dh1} is obtained by combining \eqref{eq:ddN} and \eqref{eq:diff_me},
\beq
\begin{split}
-\partial_\b \d\f \cdot\partial_\a\d\N &=
-\Brk{(\nabla_\b v^\g) - w T^\g_\b}\Brk{(\nabla_\a a_\g) - b h_{\a\g}} \\
&- \Brk{(\nabla_\b w) + v^\g h_{\b\g}}\Brk{(\nabla_\a b) + T^\d_\a a_\d}.
\end{split}
\label{eq:III}
\eeq
In remains to combine \eqref{eq:I}, \eqref{eq:II} and \eqref{eq:III} to get
\beq
\begin{split}
\d h_{\a\b} &=
h_{\a\g} (\nabla_\b v^\g) - w H_{\a\b} \\
&  - (\nabla_\a a_\b) + b h_{\a\b} \\
& -\Brk{(\nabla_\b v^\g) - w T^\g_\b}\Brk{(\nabla_\a a_\g) - b h_{\a\g}} \\
&- \Brk{(\nabla_\b w) + v^\g h_{\b\g}}\Brk{(\nabla_\a b) + T^\d_\a a_\d}
\end{split}
\label{eq:dh_gen}
\eeq

So far all the relations are exact, i.e., no assumptions have been made about the smallness of the perturbation, other than the ability to decompose it in the form \eqref{eq:df}.
Equations \eqref{eq:ab} are a set of three quadratic equations for $a_\g$, $b$, which we may solve by successive approximations.

\parag{Perturbation of a flat surface}
When the surface is flat, $h_{\a\b}=0$, \eqref{eq:ab} and \eqref{eq:dh_gen} reduce into
\[
\begin{aligned}
a_\a   &= - (\nabla_\a w) - b (\nabla_\a w) - a_\b (\nabla_\a v^\b) \\
b &= -\half (g^{-1})^{\g\b} a_\g a_\b  - \half b^2.
\end{aligned}
\]
and
\[
\d h_{\a\b} = - (\nabla_\a a_\b)  - (\nabla_\b v^\g) (\nabla_\a a_\g)  - (\nabla_\b w) (\nabla_\a b)
\]
Solving for $a_\a,b$ by successive approximation, we get
\[
\begin{aligned}
a_\a &= - (\nabla_\a w) + (\nabla_\b w)(\nabla_\a v^\b) \\
&+ \half (g^{-1})^{\b\g}(\nabla_\a w)(\nabla_\b w)(\nabla_\g w) + \O(v^2,v w^2, w^4) \\
b &= -\half (g^{-1})^{\g\b} (\nabla_\b w)(\nabla_\g w) + \O(v^2, v w, w^3)
\end{aligned}
\]
Putting it all together we obtain a simple expression for the variation of the second form,
\beq
\begin{split}
\d h_{\a\b} &= (\nabla_\a\nabla_\b w) -  (\nabla_\g w)(\nabla_\a\nabla_\b v^\g) \\
& - \half (g^{-1})^{\g\d} (\nabla_\a\nabla_\b w)(\nabla_\g w)(\nabla_\d w)  + \O(v^2,v w^2,w^4).\\
\end{split}
\label{eq:h_pert}
\eeq
We then substitute \eqref{eq:h_pert} into \eqref{eq:wswb} to calculate the variation in  bending content density,
\beq
\d w_B =
\d w_B^{(0,2)}(w)  + \d w_B^{(1,2)}(v,w) + \d w_B^{(0,4)}(w)
+ \O(v^3,v^2 w, v w^3, w^5),
\label{eq:dwB_full}
\eeq
where
\[
\begin{aligned}
\d w_B^{(0,2)}(w)  &= \frac{1}{24} \calA^{\a\b\g\d}  (\nabla_\b\nabla_\a w)  (\nabla_\d\nabla_\g w) \\
\d w_B^{(1,2)}(v,w)  &= - \frac{1}{12} \calA^{\a\b\g\d}
(\nabla_\b\nabla_\a v^\g) (\nabla_\g w) (\nabla_\g\nabla_\d w)\\
\d w_B^{(0,4)}(w) &=  - \frac{1}{24} \calA^{\a\b\g\d} (g^{-1})^{\eta\e} (\nabla_\eta w) (\nabla_\e w)
(\nabla_\b\nabla_\a w)  (\nabla_\d\nabla_\g w).
\end{aligned}
\]

\bibliographystyle{unsrt}

\begin{thebibliography}{10}

\bibitem{ESK08}
E.~Efrati, E.~Sharon, and R.~Kupferman.
\newblock Elastic theory of unconstrained {non-Euclidean} plates.
\newblock {\em J. Mech. Phys. Solids}, 57:762--775, 2009.

\bibitem{NCCC03}
U.~Nath, B.C. Crawford, R.~Carpenter, and E.~Coen.
\newblock Genetic control of surface curvature.
\newblock {\em Science}, 299:1328--1329, 2003.

\bibitem{SRS07}
E.~Sharon, B.~Roman, and H.L. Swinney.
\newblock Geometrically driven wrinkling observed in free plastic sheets and
  leaves.
\newblock {\em Phys. Rev. E}, 75:046211--7, 2007.

\bibitem{KES07}
Y.~Klein, E.~Efrati, and E.~Sharon.
\newblock Shaping of elastic sheets by prescription of {Non-Euclidean} metrics.
\newblock {\em Science}, 315:1116 -- 1120, 2007.

\bibitem{Hog85}
A.~Hoger.
\newblock On the residual stress possible in an elastic body with material
  symmetry.
\newblock {\em Arch. Rat. Mech. Anal.}, 88:271--289, 1985.

\bibitem{BG05}
M.~{Ben Amar} and A.~Goriely.
\newblock Growth and instabilities in elastic tissues.
\newblock {\em J. Mech. Phys. Solids}, 53:2284--2319, 2005.

\bibitem{Sid82}
F.~Sidoroff.
\newblock Incremental constitutive equation for large strain elasto plasticity.
\newblock {\em Int. J. Eng. Sci.}, 20:19--26, 1982.

\bibitem{SD02}
F.~Sidoroff and A.~Dogui.
\newblock Thermodynamics and duality in finite elastoplasticity.
\newblock {\em Cont. Thermomech.}, pages 389--400, 2002.

\bibitem{GN71}
A.~E. Green and P.~M. Naghdi.
\newblock Some remarks on elastic-plastic deformation at finite strain.
\newblock {\em Int. J. Eng. Sci.}, 9:1219--1229, 1971.

\bibitem{Tru52}
C.~Truesdell.
\newblock The mechanical foundations of elasticity and fluid dynamics.
\newblock {\em Indiana Uni. Math. J}, 1:125--300, 1952.

\bibitem{CG05}
P.G. Ciarlet and L.~Gratie.
\newblock A new approach to linear shell theory.
\newblock {\em Math. Models Meth. Appl. Sci.}, 15:1181--1202, 2005.

\bibitem{Cia05}
P.G. Ciarlet.
\newblock {\em An introduction to differential geometry with applications to
  elasticity}.
\newblock Springer, Dordrecht, The Netherlands, 2005.

\bibitem{Kir50}
G.~Kirchhoff.
\newblock \"uber das gleichgewicht und die bewegung einer elastischen scheibe.
\newblock {\em J. Reine Angew. Math.}, 40:51--88, 1850.

\bibitem{Lov27}
A.E.H. Love.
\newblock {\em A Treatise on the Mathematical Theory of Elasticity}.
\newblock Cambridge University Press, Cambridge, fourth edition, 1927.

\bibitem{Kar10}
T.~{von K\'arm\'an}.
\newblock Festigkeitsprobleme im maschinenbau.
\newblock In {\em {Encyclop\"adie der Mathematischen Wissenschafte}}, volume~4,
  pages 311--385. 1910.

\bibitem{Wil92}
T.~J Willmore.
\newblock A survey on willmore immersions.
\newblock In {\em Geometry and topology of submanifolds}, volume~{IV}, pages
  11--16. World Scientific, Leuven, 1992.

\bibitem{Ant76a}
S.S Antman.
\newblock Ordinary differential equations of non-linear elasticity {I}:
  foundations of the theories of non-linearly elastic rods and shells.
\newblock {\em Arch. Rat. Mech. Anal.}, 61:307--351, 1976.

\bibitem{CM04b}
P.G. Ciarlet and C.~Mardare.
\newblock An estimate of the $h^1$-norm of deformations in terms of the
  $l^1$-norm of their {Cauchy-Green} tensors.
\newblock {\em C.R. Acad. Sci. Paris, Ser. I}, 338:505--510, 2004.

\bibitem{CM04}
P.G. Ciarlet and C.~Mardare.
\newblock Recovery of a manifold with boundary and its continuity as a function
  of its metric tensor.
\newblock {\em J. Math. Pures Appl.}, 83:811--843, 2004.

\bibitem{Cia03}
P.G. Ciarlet.
\newblock The continuity of a surface as a function of its two fundamental
  forms.
\newblock {\em J. Math. Pures Appl.}, 82:253--274, 2003.

\bibitem{CC04}
P.G. Ciarlet and P.C. Ciarlet.
\newblock Another approach to linearized elasticity and korns inequality.
\newblock {\em C.R. Acad. Sci. Paris, Ser. I}, 339:307--312, 2004.

\bibitem{Koi66}
W.T. Koiter.
\newblock On the nonlinear theory of thin elastic shells.
\newblock {\em Proc. Kon. Ned. Acad. Wetensch.}, B69:1--54, 1966.

\bibitem{SMS04}
E.~Sharon, M.~Marder, and H.L. Swinney.
\newblock Leaves, flowers and garbage bags: Making waves.
\newblock {\em Amer. Sci.}, 92:254--261, 2004.

\bibitem{PS96}
E.G. Poznyak and E.V. Shikin.
\newblock Small parameters in the theory of isometric imbeddings of two
  dimensional {Riemannian} manifolds in {Euclidean} spaces.
\newblock {\em Amer. Math. Soc. Trans.}, 178:151--192, 1996.

\bibitem{GB05}
A.~Goriely and M.~{Ben Amar}.
\newblock Differential growth and instability in elastic shells.
\newblock {\em Phys. Rev. Lett.}, 94:198103--4, 2005.

\bibitem{IJ97}
G.~Iooss and D.D. Joseph.
\newblock {\em Elementary Stability and Bifurcation Theory}.
\newblock Springer, second edition, 1997.

\bibitem{Ada75}
R.A Adams.
\newblock {\em Sobolev spaces}.
\newblock Academic Press, London, 1975.

\bibitem{FW55}
Y.C. Fung and W.H. Wittrick.
\newblock A boundary layer phenomenon in the large deflection of thin plates.
\newblock {\em Quart. J. Mech. Appl. Math.}, VIII:191--210, 1955.

\bibitem{DB08}
J.~Dervaux and M.~{Ben Amar}.
\newblock Morphogenesis of growing soft tissues.
\newblock {\em Phys. Rev. Lett.}, 101:068101, 2008.

\bibitem{Far03}
E.~Farge.
\newblock Mechanical induction of twist in the {Drosophila} foregut/stomodeal
  primordium.
\newblock {\em Curr. Biol.}, 13:1365--1377, 2003.

\bibitem{HHJKUBCSBMCT08}
O.~Hamant, M.G. Heisler, H.~Jonsson, P.~Krupinski, M.~Uyttewaal, P.~Bokov,
  F.~Corson, P.~Sahlin, A.~Boudaoud, E.M. Meyerowitz, Y.~Couder, and J.~Traas.
\newblock Developmental patterning by mechanical signals in arabidopsis.
\newblock {\em Science}, 322:1650--1655, 2008.

\bibitem{LCCBLMJ08}
N.~Leblanc-Fournier, C.~Coutand, J.~Crouzet, N.~Brunel, C.~Lenne, B.~Moulia,
  and J.-L. Julien.
\newblock {Jr-ZFP2, encoding a Cys2/His2-type} transcription factor, is
  involved in the early stages of the mechano-perception pathway and
  specifically expressed in mechanically stimulated tissues in woody plants.
\newblock {\em Plant Cell. Environ.}, 31(6):715--726, 2008.

\bibitem{Shr05}
B.I. Shraiman.
\newblock Mechanical feedback as a possible regulator of tissue growth.
\newblock {\em Proc. Nat. Acad. Sci. USA}, 102:3318--3323, 2005.

\bibitem{FJM02a}
G.~Friesecke, R.D. James, and S.~M\"uller.
\newblock Rigorous derivation of nonlinear plate theory and geometric rigidity.
\newblock {\em C. R. Acad. Sci. Paris, S\'er I}, 334:173--178, 2002.

\bibitem{FJM02b}
G.~Friesecke, R.D. James, and S.~M\"uller.
\newblock A theorem on geometric rigidity and the derivation of nonlinear plate
  theory from three dimensional elasticity.
\newblock {\em Comm. Pure Appl. Math.}, 55:1461--1506, 2002.

\bibitem{FJM06}
G.~Friesecke, R.D. James, and S.~M\"uller.
\newblock A hierarchy of plate models derived from nonlinear elasticity by
  {$\Gamma$}-convergence.
\newblock {\em Arch. Rat. Mech. Anal.}, 180:183--236, 2006.

\bibitem{Dal93}
G.~{dal Maso}.
\newblock {\em An introduction to {$\Gamma$}-convergence}.
\newblock Birkhauser, 1993.

\bibitem{HH06}
Q.~Han and J.-X. Hong.
\newblock {\em Isometric enbeddings in {Riemannian} manifolds in {Euclidean}
  spaces}.
\newblock Amer. Math. Soc., 2006.

\end{thebibliography}

\end{document}